\documentclass[twocolumn,tighten]{aastex6}

\usepackage{graphicx}
\usepackage{amsmath}
\usepackage{color}
\usepackage{acronym}

\def\EAH{\textit{Einstein@Home}}
\def\Fermi{\textit{Fermi}}
\DeclareMathOperator{\sinc}{sinc}

\newcommand{\msun}{\ifmmode\mbox{M}_{\odot}\else$\mbox{M}_{\odot}$\fi}

\def\sinc{\mathrm{sinc}}

\DeclareGraphicsExtensions{.pdf}

\shorttitle{The Einstein@Home Gamma-ray Pulsar Survey}
\shortauthors{\sc Clark et al.}

\begin{document}

\title{The Einstein@Home Gamma-ray Pulsar Survey. \\ 
I. Search Methods, Sensitivity and Discovery of New Young Gamma-ray Pulsars }

\author{
C.~J.~Clark\altaffilmark{1,2,3},
J.~Wu\altaffilmark{4},
H.~J.~Pletsch\altaffilmark{1,2},
L.~Guillemot\altaffilmark{5,6,4},
B.~Allen\altaffilmark{1,7,2},
C.~Aulbert\altaffilmark{1,2},
C.~Beer\altaffilmark{1,2},
O.~Bock\altaffilmark{1,2},
A.~Cu\'ellar\altaffilmark{1,2},
H.~B.~Eggenstein\altaffilmark{1,2},
H.~Fehrmann\altaffilmark{1,2},
M.~Kramer\altaffilmark{4,8,9},
B.~Machenschalk\altaffilmark{1,2},
and L.~Nieder\altaffilmark{1,2}
}
\altaffiltext{1}{Albert-Einstein-Institut, Max-Planck-Institut f\"ur Gravitationsphysik, D-30167 Hannover, Germany}
\altaffiltext{2}{Leibniz Universit\"at Hannover, D-30167 Hannover, Germany}
\altaffiltext{3}{email: colin.clark@aei.mpg.de}
\altaffiltext{4}{Max-Planck-Institut f\"ur Radioastronomie, Auf dem H\"ugel 69, D-53121 Bonn, Germany}
\altaffiltext{5}{Laboratoire de Physique et Chimie de l'Environnement et de l'Espace -- Universit\'e d'Orl\'eans / CNRS, F-45071 Orl\'eans Cedex 02, France}
\altaffiltext{6}{Station de radioastronomie de Nan\c{c}ay, Observatoire de Paris, CNRS/INSU, F-18330 Nan\c{c}ay, France}
\altaffiltext{7}{Department of Physics, University of Wisconsin-Milwaukee, P.O. Box 413, Milwaukee, WI 53201, USA}
\altaffiltext{8}{Jodrell Bank Centre for Astrophysics, School of Physics and Astronomy, The
University of Manchester, Manchester M13 9PL, UK}
\altaffiltext{9}{University of Manchester, Manchester M13 9PL, UK}

\begin{abstract}
\noindent
We report on the results of a recent blind search survey for gamma-ray pulsars in \Fermi{} Large Area
Telescope (LAT) data being carried out on the distributed volunteer computing system, \EAH. The
survey has searched for pulsations in 118 unidentified pulsar-like sources, requiring about $10,000$
years of CPU core time. In total, this survey has resulted in the discovery of 17 new gamma-ray
pulsars, of which 13 are newly reported in this work, and an accompanying paper. These pulsars are
all young, isolated pulsars with characteristic ages between $12$~kyr and $2$~Myr, and spin-down
powers between $10^{34}$ and $4\times10^{36}$~erg~s$^{-1}$. Two of these are the slowest spinning
gamma-ray pulsars yet known. One pulsar experienced a very large glitch $\Delta f/f \approx
3.5\times10^{-6}$ during the \Fermi{} mission. In this, the first of two associated papers, we
describe the search scheme used in this survey, and estimate the sensitivity of our search to
pulsations in unidentified \Fermi-LAT sources.  One such estimate results in an upper limit of
$57\%$ for the fraction of pulsed emission from the gamma-ray source associated with the Cas A
supernova remnant, constraining the pulsed gamma-ray photon flux that can be produced by the
neutron star at its center. We also present the results of precise timing analyses for each of the
newly detected pulsars.
\end{abstract}


\newacro{LAT}[LAT]{Large Area Telescope}
\newacro{E@H}[E@H]{\textit{Einstein@Home}}
\newacro{MSP}[MSP]{millisecond pulsar}
\newacro{SSB}[SSB]{solar system barycenter}
\newacro{S/N}[S/N]{signal-to-noise ratio}
\newacro{DFT}[DFT]{discrete Fourier transform}
\newacro{FFT}[FFT]{fast Fourier transform}
\newacro{cdf}[cdf]{cumulative distribution function}
\newacro{pdf}[pdf]{probability density function}
\newacro{VHE}[VHE]{very high energy}
\newacro{PWN}[PWN]{pulsar wind nebula}
\newacro{SNR}[SNR]{supernova remnant}
\newacro{BIC}[BIC]{Bayesian Information Criterion}
\newacro{AIMC}[AIMC]{Affine Invariant Monte~Carlo}
\newacro{DM}[DM]{dispersion measure}

\section{Introduction}\label{s:intro}
Since the launch of the \textit{Fermi Gamma-ray Space Telescope} in 2008, the on-board Large Area
Telescope \citep[LAT;][]{generalfermilatref} has increased the number of known gamma-ray pulsars
from around 10 to over 200.\footnote{\url{http://tinyurl.com/fermipulsars}} Indeed, thanks to the
LAT, we now know pulsars to be the dominant individual gamma-ray source class within the Milky Way
galaxy \citep[Second Fermi Large Area Telescope Catalog of Gamma-ray Pulsars, hereafter
  2PC,][]{2PC+2013}.

Two-thirds of gamma-ray pulsars were first detected by observations in other wavelength regimes
(e.g. radio or X-ray pulsars), the rotation ephemerides from which could then be used to ``phase
fold'' the LAT photon arrival times to test for pulsed gamma-ray emission. However, approximately
one third of the LAT-detected pulsars were unknown prior to the discovery of pulsations in their
gamma-ray flux
\citep{Abdo2008+CTA1,Abdo2009+16BSPs,SazParkinson2010+8BSPs,Pletsch+2012-9pulsars,Pletsch2012+J1838,Pletsch+2012-J1311,Pletsch+2013-4pulsars}. Only
a handful of these pulsars were subsequently detected in radio observations, the others could not
have been discovered without ``blind'' searches in gamma-ray data.

The recent \textit{Fermi}-LAT Third Source Catalog \citep[3FGL;][]{3FGL} includes 3033 gamma-ray
sources, of which about $1000$ lack strong associations or likely counterparts from observations at
other wavelengths. Many of these sources have similar properties to the known gamma-ray pulsars
(i.e. low time variability and a highly curved spectrum). A large effort is underway to identify
pulsars amongst the unidentified pulsar-like gamma-ray sources, both by performing dedicated radio
searches targeting the locations of LAT sources
\citep[e.g.][]{Camilo2015+ParkesMSPs,Cromartie2016+AreciboMSPs}, and by searching amongst the LAT
data itself for pulsations. This paper, and an accompanying paper (J.~Wu et al., 2017, in preparation,
hereafter Paper II), will describe the latest results from the latter technique.

Due to the sparsity of the LAT photon data (only about $10$ photons per day are detected from a
typical gamma-ray pulsar), blindly searching for pulsations amongst the LAT photon arrival times is
an enormously expensive computational task. For weak pulsar signals, long integration times are
required to reach a detectable \ac{S/N}, and, as a result, signal parameters must be searched
with an extremely fine resolution to avoid losing the signal.

In addition, ensuring sensitivity to the more extreme gamma-ray pulsars, such as young pulsars with
high spin-down rates or faster spinning \acp{MSP}, requires searching over vast parameter volumes,
and therefore incurs a proportionally large computation cost. To meet these requirements, we utilize
the computing power of the \EAH{} project, which distributes the computations amongst the many
thousands of participating volunteers' devices \citep{Allen2013+EatH}.  We began performing blind
searches for gamma-ray pulsars on \EAH{} in 2011, with the first such survey resulting in the
detection of four new pulsars \citep{Pletsch+2013-4pulsars}.

In a recent study \citep{Methods2014}, we investigated the efficiency of different blind search
schemes \citep[e.g.,][]{Atwood2006}, and developed new techniques to boost the sensitivity of a
search without impacting its computational cost. These new methods are currently being used as part
of a large-scale survey of pulsar-like \textit{Fermi}-LAT sources, running on \EAH. In
combination with the recent ``Pass 8'' improvements to the LAT event reconstruction
\citep{Pass8}, these advances have had a spectacular effect on the blind search sensitivity, an
early indication of which was given by the detection of PSR~J1906$+$0722 within one of the first
sources searched in this survey \citep{Clark2015+J1906}.

In this paper, we present $13$ new pulsar discoveries from the full \EAH{} survey of $118$
sources. These are the result of around $10,000$~years of CPU time generously donated by volunteers.

The paper is organized as follows. In Section \ref{s:scheme} the search methods are described; an
investigation of the sensitivity of the search follows in Section \ref{s:sensitivity}; details of
the newly discovered pulsars and their timing solutions are given in Section \ref{s:timing};
Section \ref{s:discussion} contains a discussion of the sensitivity of blind searches to
unidentified gamma-ray pulsars; and finally we summarize our conclusions in Section \ref{s:concl}.

In Paper II, we will describe in more detail the LAT data preparation procedures; the selection of
target sources and ranking based on their spectral properties; the identification of candidate
multiwavelength counterparts; phase-resolved gamma-ray spectral analyses of the newly discovered
pulsars; and the results of dedicated follow-up radio pulsation searches.

\section{Search Scheme}\label{s:scheme}
\subsection{Data}\label{s:data}
The data searched during the survey consisted of gamma-ray photons detected by the LAT between
2008~August~4 and 2014~April~6 (2014~October~1 for some sources searched later in the survey) with
energies above $100$~MeV. Photons were included if they arrived within $8\arcdeg$ of a target
source, with a zenith angle $<100\arcdeg$ and when the LAT's rocking angle was $<52\arcdeg$.  The
photons were selected and analyzed using the \texttt{P8\_SOURCE\_V3} instrument response functions
(IRFs).

For each target source, we performed a likelihood spectral analysis using the \texttt{pointlike}
package \citep{Kerr2010+Pointlike}. Our source model included all 3FGL catalog sources within
$13\arcdeg$ of the target source and used the \texttt{template\_4years\_P8\_V2\_scaled.fits} map
cube and \texttt{isotropic\_source\_4years\_P8V3} template to model the Galactic diffuse emission
\citep{Acero2016+Diffuse} and isotropic background respectively.

Target sources were modeled with an exponentially cutoff power law typical of gamma-ray pulsars.
During the likelihood fitting, we allowed the normalization of the diffuse models, and the spectral
parameters of the target source and all 3FGL sources within $5\arcdeg$ to vary. Sources searched
near the beginning of the survey had their sky positions fixed at the 3FGL location. Later sources
were relocalized during the likelihood fitting to exploit the improved angular resolution offered by
the Pass 8 data. Spectral energy distribution (SED) plots and Test Statistic (TS) maps were visually
compared to the corresponding 3FGL sources to diagnose any problems with the fitting. With the
best-fitting source model, we used \texttt{gtsrcprob}\footnote{\texttt{gtsrcprob} is part of the
  \textit{Fermi} Science Tools, available at
  \url{http://fermi.gsfc.nasa.gov/ssc/data/analysis/software/}} to compute weights representing
the probability of each photon having come from our target source based on their reconstructed
energy and arrival direction. Full details of the data preparation methods, and a description of how
target sources were prioritized for searching, will be given in Paper~II.

The IRFs and diffuse templates used here were internal pre-release versions of the Pass 8 analysis
tools because the final release versions were not yet available when the survey began. When
investigating the gamma-ray emission from the region surrounding PSR~J1906$+$0722
\citep{Clark2015+J1906}, we found that these preliminary IRFs and templates resulted in spectral
parameters consistent with those found using the final Pass 8 release. However, photon weights
calculated with the most recent Pass 8 data usually result in slightly higher pulsation significance
within the same time interval; the sensitivity estimates in Section \ref{s:sensitivity} are likely
to be more conservative as a result.

\pagebreak
\subsection{Parameter Space}\label{s:param_space}
To search for gamma-ray pulsations in LAT data, it is necessary to assume a certain ``phase model''
(i.e. a rotation ephemeris) relating the arrival time of every photon to a certain rotational phase,
and test all possible combinations of the model parameters for pulsations, indicated by large values
of a detection statistic (described in Section \ref{s:det_stats}). In the case where a signal is
present, the distribution of rotational phases will deviate significantly from uniformity. For
isolated pulsars, the phase model\footnote{While we define the phase in radians, in all plots we
  show phase in rotations for clarity, and re-normalize the pulse profiles accordingly.} is
typically described by a Taylor series expansion in time around a chosen reference epoch $t_{\rm
  ref}$, for photon arrival time $t$ at the \ac{SSB},
\begin{equation}
  \Phi(t) = \Phi_0 + 2\pi\,\sum_{m=1}\frac{f^{(m-1)}}{m!}(t-t_{\rm ref})^m\,,
  \label{e:phase_model}
\end{equation}
where $f^{(m)}$ denotes the $m$th time derivative of the pulsar's rotational frequency, $f$. While
the higher derivative terms are often measurable for young pulsars, it is usually sufficient (and
often only feasible) to include only the first two terms in the blind search, resulting in a
simplified phase model in which the spin frequency decreases by a constant spin-down rate, $\dot{f}
\equiv f^{(1)}$.

Aside from correcting for this constant spin-down, it is also necessary to account for the apparent
Doppler modulation of pulsations that results from the Earth's orbit around the \ac{SSB}. This can
be achieved by applying position-dependent corrections to the measured photon arrival times, to
retrieve the set of arrival times at the \ac{SSB}, hereafter denoted as $\left\{t_j\right\}$. The
angular resolution at which sky positions must be searched increases linearly with the pulsar's spin
frequency (see Equation \ref{e:sky_spacing}). For all but the slowest of pulsars, the required
resolution is finer than the gamma-ray source localization, determined by the LAT's point-spread
function. For a blind survey of unidentified gamma-ray point sources, it is therefore necessary to
search in two sky positional parameters (R.A. $\alpha$ and decl. $\delta$), making the overall search
parameter space four-dimensional. For sources for which we used the original 3FGL locations, we
searched a circular region around the source with an angular radius that was 50\% larger than the
semi-major axis of the 95\% confidence region. For relocalized sources, we searched a conservatively
large region with a radius three times larger than the semi-major axis of the $68\%$ confidence
region.

Some pulsars (e.g. PSRs~J2017$+$3625 and J1350$-$6225) were found near the edge, or even slightly
outside of their search regions, indicating that the confidence regions may be underestimated, and
pulsars may have been missed by our survey as a result. This could be due to nearby, unmodeled
gamma-ray sources ``pulling'' the apparent position of the source away from its true position, as
was seen with PSR~J1906$+$0722 \citep{Clark2015+J1906}. To mitigate this effect in future surveys it
may be necessary to search over larger regions, especially for sources at low Galactic latitude,
where source confusion is more likely. However, increasing the solid angle over which we search
increases the computational cost of the search by the same factor.

\begin{figure*}
	\centering
	\includegraphics[width=0.72\textwidth]{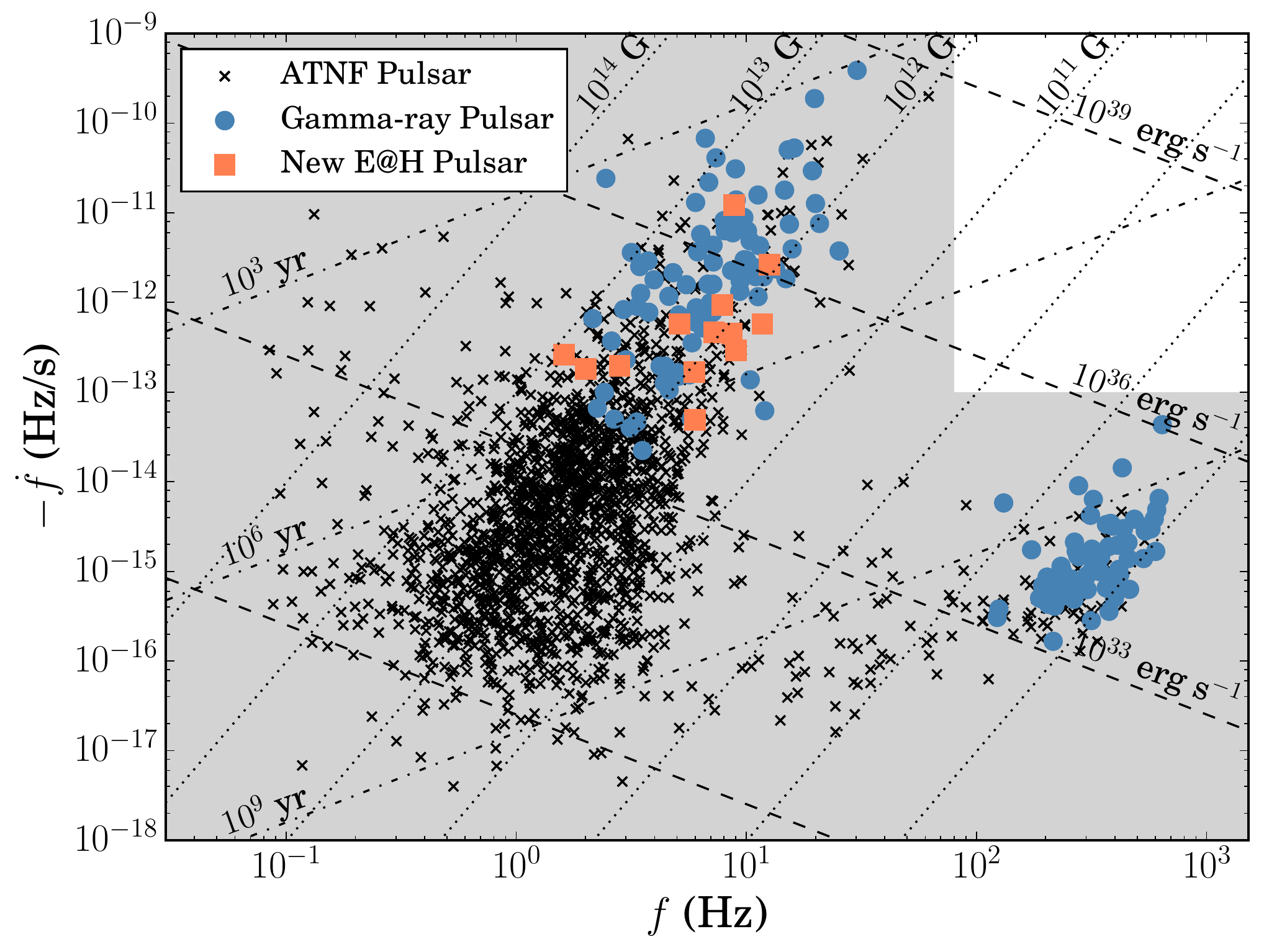}
	\caption{Frequency--spin-down diagram, showing the locations of non-gamma-ray pulsars in the ATNF
    Pulsar Catalogue \citep[black crosses,][]{ATNF2005}, gamma-ray pulsars detected by \Fermi-LAT
    (blue circles), and the newly detected \EAH{} pulsars reported in this work (orange
    squares). The parameter space covered by the \EAH{} survey is shown by the gray shaded
    area. Lines of constant characteristic age, $\tau_{\rm c} = -f/2\dot{f}$ (dotted-dashed), surface
    magnetic field strength, $B_{\rm S} = (- 1.5 I c^3 \dot{f} f^{-3})^{1/2}/2\pi R_{\rm S}^3$,
    (dotted) and spin-down power, $\dot{E} = - 4\pi^2 I f \dot{f}$ (dashed) are also shown. To
    calculate these, we assumed neutron star moments of inertia, $I = 10^{45}$\,g~cm$^2$ and radii,
    $R_{\rm S} = 10$\,km, as in e.g. \citet{2PC+2013}. The timing analyses performed on the newly
    discovered pulsars, the results from which were used to calculate the properties above, are
    described in Section~\ref{s:timing}.}
	\label{f:f_fdot_diagram}
\end{figure*}

We split the search parameter space into two main regions: the young pulsar region, with spin
frequencies below $80$~Hz; and the \ac{MSP} region at higher spin frequencies.  This parameter space
is shown by the shaded area in Figure \ref{f:f_fdot_diagram}, and covers all currently known young
pulsars, \ac{MSP}s and magnetars.  In the low-frequency region we extend the $\dot{f}$ range from
$0$ down to $-10^{-9}$~Hz~s$^{-1}$ to be sensitive to the youngest and most energetic
pulsars. Older, recycled \ac{MSP}s have much lower spin-down rates, and we therefore only search
from $0$ down to $-10^{-13}$~Hz~s$^{-1}$ in this region.  Since more sky locations must be searched
at higher frequencies, the majority of the computational cost of the search is spent in the
high-frequency and high-spin-down regions. Pulsars whose pulse profile features two similarly sized
peaks separated by half a rotation have most power in the second harmonic of their spin
frequency. For this reason, we search up to $1520$~Hz, more than twice the frequency of the fastest
known \ac{MSP}, $716$~Hz \citep{Hessels2006+FastestMSP}. Only one known pulsar, PSR~J0537$-$6910,
has its second spin harmonic outside our parameter space \citep{Marshall1998+J0537}.

For each of the 118 unidentified LAT sources in which we searched, this parameter space is split
into $\sim10^3$--$10^6$ smaller \textit{work units}, each of which can be searched in a few hours on a
typical home computer. These work units are then distributed amongst \EAH{} volunteers' computers.

\subsection{Detection Statistics}\label{s:det_stats}
In all stages of a gamma-ray pulsar search, statistical tests are used to measure the strength of
pulsations for given rotational parameters. The detection statistics used in this survey are
described in detail in \citet{Methods2014}, and briefly defined here.

\citet{Kerr2011} demonstrated the advantages of applying a weight to each photon indicating its
probability of having come from the target source. The photon probability weights mentioned in
Section \ref{s:data}, denoted by $\left\{w_j\right\}$, were therefore used to weight the
contributions of each photon to a detection statistic. Weighting photons improves the sensitivity of
a blind search by avoiding the need to apply specific photon energy and angular offset cuts, and by
increasing the apparent fraction of flux that is pulsed.

To mitigate the computational cost of a blind search, semicoherent methods can be used, in
which only photons arriving within a certain time difference from one another are combined
coherently. The (real-valued) semicoherent detection statistic used in this search is defined as\footnote{The
  subscript $1$ here denotes that the detection statistic only sums power in the fundamental
  harmonic.}
\begin{equation}
  S_1 = \frac{1}{\kappa_{S_1}} \sum_{j=1}^N \sum_{k\neq j}^{N} w_j w_k \, e^{-i\left[\Phi(t_j) - \Phi(t_k)\right]} \,
  \hat{W}_T^{\rm\tiny{rect}}(\tau_{jk})\,,
  \label{e:s1_definition}
\end{equation}
where $\tau_{jk}$ is the time difference, or lag, between the arrivals of the $j$th and $k$th
photons, and $\hat{W}_T^{\rm{rect}}$ is a rectangular window, of length $T$,
\begin{equation}
  \hat{W}_T^{{\rm\tiny rect}}(\tau) =
      \begin{cases}
	1, & |\tau| \leq T/2 \\
	0, & \text{otherwise}  \,.
 	\end{cases}
	\label{e:rectlagwin}
\end{equation}
The $\kappa_{S_1}$ term of Equation (\ref{e:s1_definition}) is a normalizing factor,
\begin{equation}
  \kappa_{S_1} = \sqrt{\sum_{j=1}^N \sum_{k\neq j}^{N} w_j^2 w_k^2
    \hat{W}_T^{\rm\tiny{rect}}(\tau_{jk})}\,,
  \label{e:kappa_S_1}
\end{equation}
making the noise distribution of $S_1$ well approximated by a normal distribution with zero mean and
unit variance.

The lag window length, $T$, is an important tuneable parameter for a search based on
semicoherent methods. A longer lag window offers more sensitivity, but requires a finer grid in all
four search parameters, and therefore results in a more costly search.

In the case where the lag window covers the entire observation span, then all photons are combined
fully coherently, and the test statistic reduces to the well-known Rayleigh test (modulo a constant term,
and normalization) at the fundamental harmonic, $n = 1$,
\begin{equation}
	\mathcal{P}_n = \frac{1}{\kappa^2}
	\left|\sum_{j=1}^N w_j \, e^{-i n \Phi(t_j)}\right|^2\,,
	\label{e:Pndef}
\end{equation}
with the normalization constant,
\begin{equation}
	\kappa^2 = \frac{1}{2} \sum_{j=1}^N w_j^2\,.
  \label{e:kappa}
\end{equation}
This is hereafter referred to as the coherent Fourier power at the $n$th harmonic.

To gain further sensitivity to weak signals, one can also combine the coherent Fourier power from
several harmonics of the fundamental spin frequency. The well-known $H$-test developed by
\citet{deJager+1989} offers a heuristic method for combining these harmonics in the typical case
where the pulsar's pulse profile (and hence the distribution of Fourier power amongst the different
harmonics) is unknown in advance by maximizing over the number of included harmonics, $M$, via
\begin{equation}
	H = \max_{1 \leq M \leq 20}
	\left(\sum_{n=1}^M \mathcal{P}_n - 4M + 4\right)\,.
  \label{e:Htest}
\end{equation}
Combining Fourier power from higher harmonics requires finer resolution in all phase model
parameters. It therefore only becomes feasible in later search stages, in which the parameter space
within which a candidate signal could lie is constrained to be very narrow.

As discussed in \citet{Methods2014}, a multistage search scheme can be used to combine the
efficiency of a semicoherent search with the superior sensitivity of fully coherent methods. In
this scheme, the majority of the search is spent scanning the entire parameter space with the most
efficient method available, before ``following-up'' the most interesting candidates in more
sensitive stages.

\subsection{Initial Search Stage}\label{s:semicoh_stage}
In this survey, the first stage used the semicoherent detection statistic, $S_1$, with a lag window of
length $T = 2^{21}$s $\approx 24$ days. This lag window is a factor of two longer than in previous
\EAH{} searches \citep{Pletsch+2013-4pulsars}.

As described in \citet{Methods2014}, the semicoherent detection statistic, $S_1$, defined in
Equation~(\ref{e:s1_definition}), can be approximated more efficiently as a \ac{DFT}, by utilizing
the FFTW \ac{FFT} algorithms \citep{FFTW05}. We hereby refer to the DFT form of the semicoherent
detection statistic as $\hat{S}_1$.

Each FFT searched over a frequency bandwidth of $\Delta f_{\rm BW} = 32$~Hz. We applied the
technique of complex heterodyning, i.e. multiplying the FFT input vector by an additional sine wave
at the heterodyning frequency, $f_{\rm H}$, to shift the search band to higher frequencies,
$\left[f_{\rm H} - 16, f_{\rm H} + 16\right)$~Hz, without increasing the FFT memory size, $\Delta
  f_{\rm BW} T = 256$~MiB. This allows us to search for high-frequency signals, such as those from
  MSPs, using typical computing hardware. Furthermore, since the required resolution
  in the sky position becomes finer at higher frequencies, we can construct individual sky grids for
  each frequency band to avoid oversampling sky positions at low frequencies. The first frequency
  band was centered at $0$~Hz, and all frequencies below $5$~mHz were ignored to prevent harmonics of
  \Fermi's orbital frequency ($\sim0.175$~mHz) from ``drowning out'' any astrophysical signal.

To ensure approximately equal sensitivity throughout the frequency band, we performed lag-domain
interpolation \citep{Methods2014}, whereby each photon pair is interpolated into the $15$ nearest
bins on either side in the lag-series using a Welch-windowed sinc kernel
\citep[][p. 176]{Welch1967,LyonsDSP}. Since this technique introduces an additional computational
cost per pair of photons, we performed a photon weight cutoff to include at most the $30,000$
highest-weight photons, ensuring that the \ac{FFT} computation time remained the dominant
factor. Identifying the photon weights as the probability of each photon being from a pulsar, the
maximum (coherent) \ac{S/N} is proportional to $\sum_{j=1}^N w_j^2$. For sources where fewer than
$30,000$ photons were required to reach $95\%$ of this maximum (typically sources far from the
Galactic plane, where the diffuse background is lower) we increased the number of interpolation
bins, up to a maximum of $30$.

A signal whose parameters, denoted by the vector $\boldsymbol{u}$, lie within the search space will,
in general, not lie exactly at one of our search-grid points, and some of the S/N is lost as a result
of this offset, $\Delta \boldsymbol{u}$. We call this (fractional) loss in S/N \textit{mismatch},
\begin{equation}
  m = 1 - \frac{\hat{S}_1\left(\boldsymbol{u} + \Delta
    \boldsymbol{u}\right)}{\hat{S}_1\left(\boldsymbol{u}\right)}\,.
\label{e:mismatch}
\end{equation}
We can predict the expected mismatch as a function of the distance to the nearest search-grid point
using an analytical ``metric'' approximation, as described in \citet{Methods2014}. This prediction
can then be used to construct an efficiently spaced grid of points in the parameter space at which
to test for pulsations.

The spacing of frequency trials is fixed by the DFT formulation of $\hat{S}_1$ to be
\begin{equation}
  \Delta f = \frac{1}{T}\,.
  \label{e:spacing_f}
\end{equation}
While this spacing would result in a large average mismatch, we can improve upon this by performing
simple ``interbinning'' \citep{VanderKlis1989, Astone2010} to partially recover the lost S/N
experienced by signals lying between our frequency bins. This technique does not recover the full
S/N for such signals, but is far more efficient than the alternative of ``zero-padding'' the FFT to
double length.

In the remaining parameters, we construct a cubic lattice with spacings chosen to provide the optimal
maximum mismatch in each parameter of $m = 0.15$ according to the metric approximation. In $\dot{f}$
the spacing depends on the lag window~$T$ but also requires a refinement based on the full data set \citep{PletschAllen2009},
\begin{equation}
  \Delta \dot{f} = \frac{12\sqrt{10 m}}{\pi T^2}\left[1 + \frac{60}{N}\sum_{j=1}^N\frac{(t_j
      - t_{\rm ref})^2}{T^2}\right]^{-1/2}\,.
\end{equation}
The grid of sky locations is first defined within a circle (with unit radius) in the ecliptic plane
as a square grid with spacings in each direction of
\begin{equation}
    \Delta n_{x} = \Delta n_{y} = \frac{2 \sqrt{m}}{\pi f_{\rm max} r_{\rm E}} \left[1 -
      \sinc^2(\Omega_{\rm E} T/2)\right]^{-1/2}\,,
    \label{e:sky_spacing}
\end{equation}
where $f_{\rm max}$ is the maximum frequency in the searched frequency band, $r_{\rm E}$ and
$\Omega_{\rm E}$ are the Earth's orbital semi-major axis (in light seconds) and orbital angular
frequency respectively, and $\sinc(z)=\sin(z)/z$. These locations are then projected back into the
celestial sphere to cover the LAT source localization region. At each location, barycentering
corrections are applied to each photon's arrival time according to the JPL DE405 solar system
ephemeris.

Each work unit performs an FFT at every location in this cubic lattice within its assigned portion of
the parameter space. The five highest values of $\hat{S}_1$ (including interbinned samples) are
stored in a running short list that is updated after each \ac{FFT}. At the end of the semicoherent
stage, this short list is saved, and each short-listed candidate is automatically ``followed up'' in
additional, more sensitive search stages.

\subsection{Follow-up and Refinement Stages}\label{s:scheme_followup}
After the semicoherent stage, we are left with a small number of candidates in each work unit that
have been localized to a small region of the parameter space. However, due to the large number of
work units for each \Fermi-LAT source, weak signals in these short lists can be of low overall
significance. To separate weak signals from noise candidates, we can carry out more sensitive
follow-up stages to act as a veto for the large number of candidate signals reported back by the
semicoherent stage.

In the \EAH{} survey, we implemented an intermediate refinement stage, in which candidates from the
first stage are refined using a double-length lag window ($T = 2^{22}$ s $\approx 48$ days). This
step is computationally cheap, and narrows down the volume in which the candidate signal lies by a
factor of $\sim16$.

Following the semicoherent refinement stage, we now have a parameter space volume around each
candidate that is small enough for a fully coherent search to be feasible with just a small
associated computing cost. For this stage, we search only in the fundamental harmonic using the
$\mathcal{P}_1$ test, with grid spacings according to the coherent metric approximation derived in \citet{Methods2014}.

All search stages up to this point are carried out on the \EAH{} volunteers' computers, after which
the short-listed candidates from the initial semicoherent
stage (each of which were followed up), and the top candidates from the coherent follow-up stage
are sent back to our servers.

As results are sent back, we update the top 20 most significant coherent candidates (see Appendix
\ref{a:PFA_ranking} for a description of the ranking procedure) overall from each source, and
perform further follow-up and refinement procedures on them. First, we refine the location of the
candidate using the $\mathcal{P}_1$ statistic, but calculated over a grid with a smaller mismatch
($m = 0.05$) than that used in the third stage. We then perform a fully coherent search using the
$H$-test to incoherently sum the Fourier power in the first five harmonics.

After this refinement step, diagnostic plots for each candidate are produced that illustrate the
candidates' signals and their evolution throughout the \Fermi-LAT observation time. This allows us
to identify pulsars with timing noise, whose pulsations may be visible in these plots despite having
a low apparent coherent power due to variations in their signal phase.

\section{Sensitivity}\label{s:sensitivity}
In \citet{Methods2014}, the sensitivity of a blind search for gamma-ray pulsations was defined as the
minimum pulsed fraction of the observed photon flux that can be detected with a
fixed probability, $P_{\rm det}^{\ast}$, and with a fixed false-alarm probability, $P_{\rm FA}^{\ast}$. We now
apply this definition to investigate the sensitivity of our search to each source in the survey.

\begin{figure}
	\centering
	\includegraphics[width=\columnwidth]{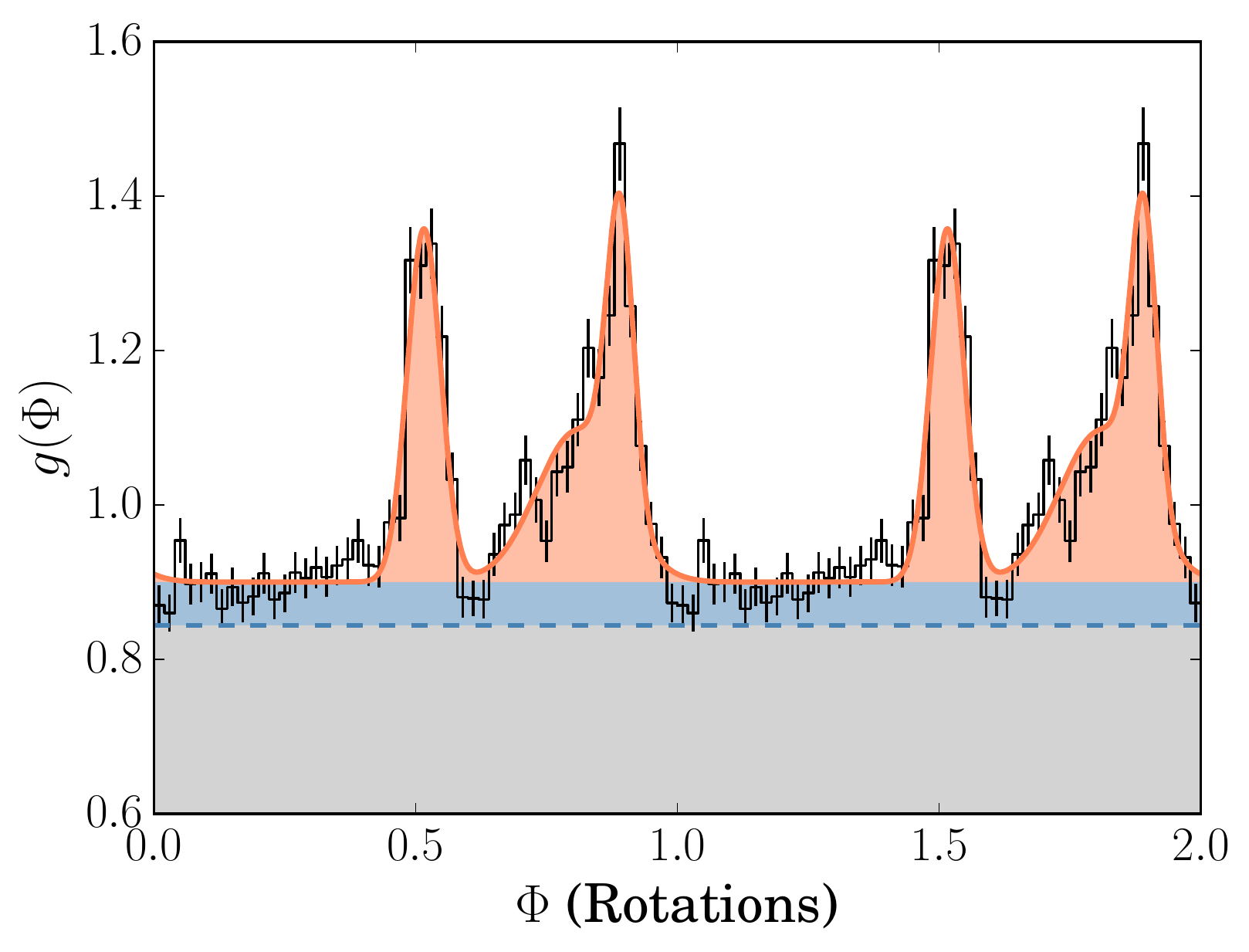}
	\caption{Gamma-ray pulse profile of PSR~J2017$+$3625 illustrating the definition of the pulsed
    fraction. The blue dashed line indicates the background level, $b$. The pulsed fraction is
    defined by the area under the template pulse profile above its lowest level (the orange
    shaded area), divided by the source fraction, $s$ (that is the sum of blue and orange areas).}
	\label{f:pulse_profile_example}
\end{figure}

The quantity of interest is the fraction of the background-subtracted weighted photon flux that is
pulsed, denoted $p_s$. Given a set of photon weights, the fractions of the weighted photon counts
that can be attributed to the source, $s$, and background fraction $b$, are estimated as
\citep{Guillemot2012,2PC+2013}
\begin{align}
  s = \frac{\sum_{j=1}^N w_j^2}{\sum_{j=1}^N w_j}\,,\qquad
  b = 1 - s\,.
\end{align}
The probability of the $j$th photon being pulsed is $w_j \, p_s$, and the overall
weighted pulse profile takes the form
\begin{equation}
  g(\Phi) = \frac{b}{2\pi} + s\, g_s\left(\Phi\right)\,,
\end{equation}
where $g_s(\Phi)$ is the background-subtracted pulse profile,
\begin{equation}
  g_s(\Phi) = \frac{1 - p_s}{2\pi} + p_s\, g_p\left(\Phi\right)\,,
\end{equation}
where $g_p(\Phi)$ is the pulse profile after subtracting all unpulsed emission (background or
otherwise). These quantities are illustrated in Figure \ref{f:pulse_profile_example}.

Note that this definition of the pulsed fraction is equal to the area under the pulse in the
normalized pulse profile, as opposed to the ``rms pulsed flux'' used by e.g. \citet{Dib2009+RMS},
which is additionally dependent on the shape of the pulse profile. While the rms pulsed flux
provides a measure of the power of pulsations, this does not provide a physically meaningful measure
of the proportion of pulsed flux emitted by the pulsar \citep{Zhu2008+PulsedFraction}.

\begin{figure}
	\centering
	\includegraphics[width=0.95\columnwidth]{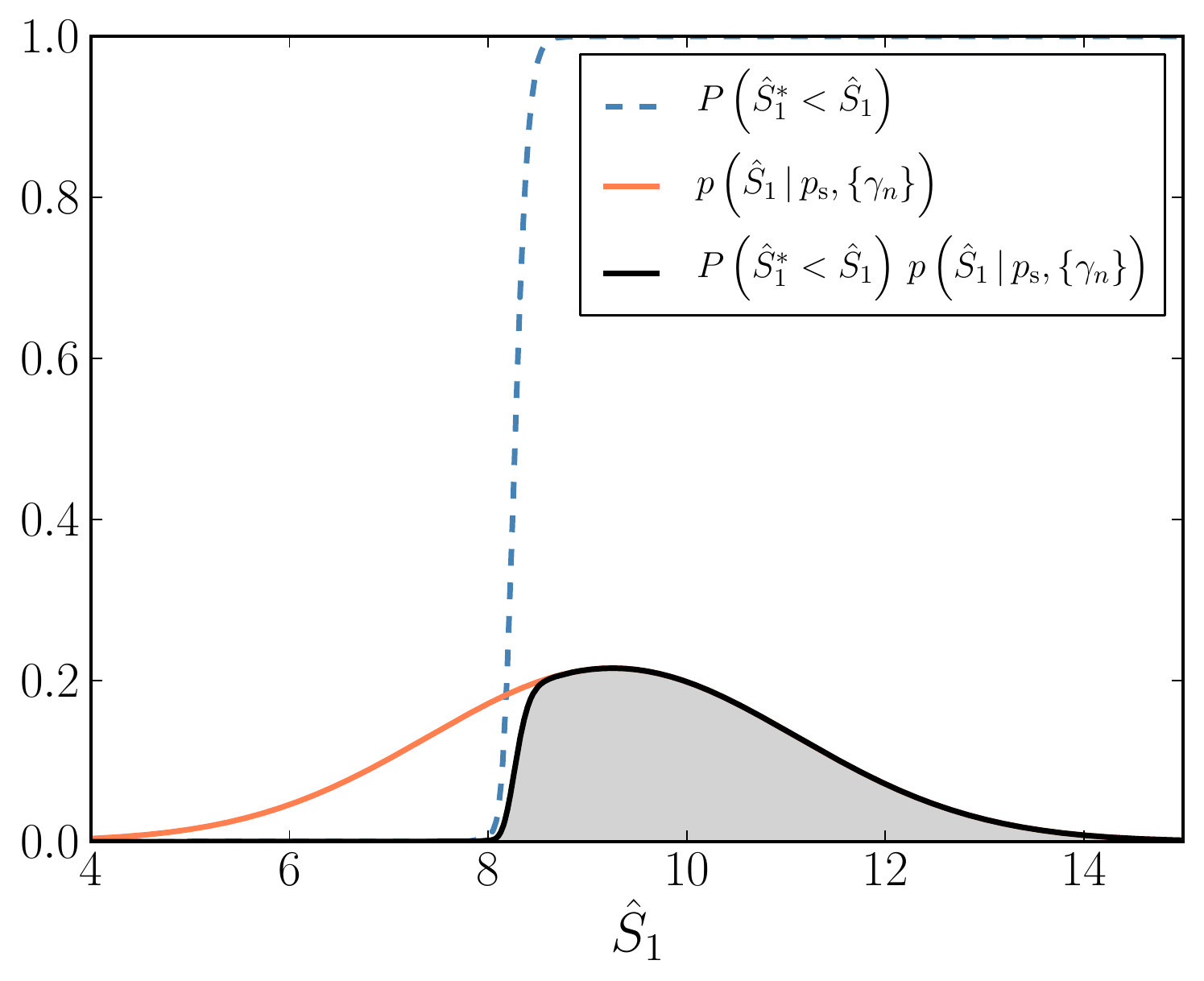}
	\caption{Illustration of the definition of the detection probability. The orange line
    indicates the \ac{pdf} of the semicoherent detection statistic (including mismatch) in the
    presence of a signal with fixed pulse profile and pulsed fraction. The blue dashed line
    shows the empirical probability that a true signal resulting in a detection statistic $\hat{S}_1$
    will be followed-up (and hence detected). The detection probability is therefore the area under
    the product of these functions, shown by the grey shaded area.}
	\label{f:detection_probability}
\end{figure}

For a pulsar to be detected by this survey, its signal must be strong enough to enter the
short list of semicoherent candidates within the work unit that covers the region of the parameter
space in which the signal lies. That is, the measured value of $\hat{S}_1$ at the grid point nearest
the signal's location in the parameter space must be greater than the lowest value in the short list
of candidates, $\hat{S}_1^{\ast}$. The probability that a signal with a pulse profile described by
the complex Fourier coefficients $\left\{\gamma_n\right\}$ will be detected by the survey, as
a function of the pulsed fraction is
\begin{equation}
  P_{\rm det} \left(p_s \,| \left\{\gamma_n\right\}\right) = \int_{-\infty}^{\infty} P\left(\hat{S}^{\ast}_1 <
  \hat{S}_1\right) p \left(\hat{S}_1 \, | \, p_{\rm s}, \left\{\gamma_n\right\}\right) d\hat{S}_1 \,,
  \label{e:pdet}
\end{equation}
where $P\left(\hat{S}^{\ast}_1 < \hat{S}_1\right)$ is the (empirically measured) probability that
$\hat{S}_1^{\ast}$ is less than $\hat{S}_1$, and $p \left(\hat{S}_1 \, | \, p_{\rm s},
\left\{\gamma_n\right\}\right)$ is the probability density function of the measured value of
$\hat{S}_1$ for a signal at a random location within the searched parameter space (see Appendix
\ref{a:app_signal_stats} for the derivation of this distribution). Each of these quantities depends
additionally on the set of photon weights for each source; we have omitted these dependencies from
Equation (\ref{e:pdet}) for readability. This definition of the detection probability is illustrated
in Figure \ref{f:detection_probability}.  This equation can be solved numerically to recover the
minimum pulsed fraction, $p^{\ast}_{\rm s}$, that can be detected at a given probability.

The purpose of the coherent follow-up stage of the survey is to greatly improve the significance of
any true signal that is detected by the first stage, and we apply our final false-alarm probability
threshold to the candidates from this stage. For typical values of $p_s$ given by solving Equation
(\ref{e:pdet}), the expected coherent power corresponds to an extremely significant signal
($\mathcal{P}_1 \approx 270$, $P_{\rm FA} \sim 10^{-59}$) and hence even a conservative false-alarm
threshold has no real effect on the overall sensitivity estimate. In practice, effects such as
glitches or timing noise that are not included in our simplified isolated pulsar phase model can
severely reduce the observed coherent power, resulting in true signals with large $p_s$ appearing
with low significance. We attempt to mitigate these effects somewhat by monitoring the 20 most
significant candidates from both the semicoherent stage and the follow-up stages rather than
applying the false-alarm threshold rigorously.

\begin{figure}
    \centering
    \includegraphics[width=\columnwidth]{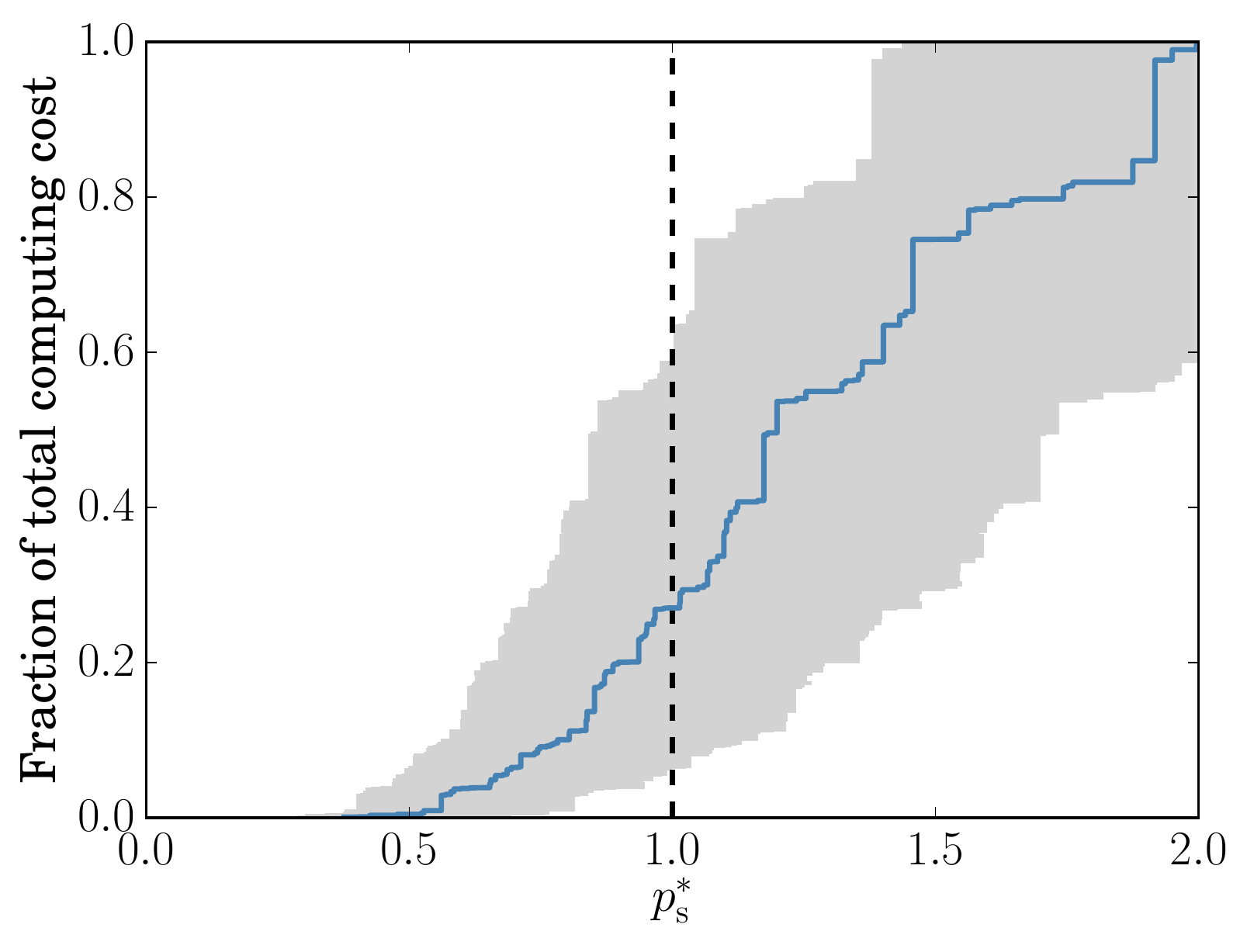}
    \caption{Cumulative fraction of the total computing cost of the survey, as a function of the
      estimated pulsed fraction threshold. The solid line shows the 95\% detection probability
      pulsed fraction threshold averaged over the pulse profiles of the 30 pulsars in the 2PC with
      the highest significance. The shaded area shows the range of pulsed fraction thresholds over
      these pulse profiles. The dashed vertical line denotes the maximum expected pulsed fraction of
      $p_{\rm s}^{\ast} = 1$.}
  \label{f:ps_nffts}
\end{figure}

In Table \ref{t:minimum_pulsed_fraction} we estimate the minimum pulsed fraction, $p_s^{\ast}$, that
can be detected with $P_{\rm det}^{\ast} = 0.95$ for each source in our survey, averaged over the
pulse profiles from the 30 most significant pulsars in the 2PC, and assuming constant signal
parameters (i.e. no glitches or significant timing noise). For sources in which a new gamma-ray
pulsar was discovered, we also report the measured pulsed fraction for illustration. Note that while
some sources have $p_s^{\ast} > 1$, this does not necessarily preclude the detection of pulsed
emission from this source, since the assumptions on which these estimates are based might not always
apply. For example, the true pulse profile could be narrower than average or the photon weights
might not accurately represent the probability of each photon coming from the target source.  The
results of this sensitivity estimation displayed in Figure \ref{f:ps_nffts} therefore also
illustrate the range of thresholds covered by the various pulse profiles observed.

While the semicoherent search stage is rather robust to the presence of timing noise, any large
enough deviation from our simplified constant spin-down model will significantly affect the
sensitivity of our search. Glitching pulsars, for example, are particularly difficult to detect
because their large jumps in spin frequency prevent the S/N from accumulating steadily throughout
the observations. Pulsars in binary systems are all but impossible to find in a search for isolated
pulsars. Our pulsed fraction thresholds therefore only represent our sensitivity to well-behaved
isolated pulsars. In particular, our sensitivity estimates are likely to be most reliable for
sources far from the Galactic plane, where we would expect to find old, stable MSPs. Our sensitivity
estimates rely on the assumption that a pulsar lies within the sky region in which we search. As
mentioned in Section \ref{s:param_space}, in some cases, the extension of this region may have been
underestimated.

\floattable
\begin{deluxetable}{cccccccccc}
  \tabletypesize{\scriptsize}
  \tablecaption{\label{t:minimum_pulsed_fraction} Pulsed fraction upper limits}
  \tablecolumns{10}
  \tablehead{
    \colhead{3FGL Name} & \colhead{Pulsar Name} & \colhead{Max.\tablenotemark{a}\; $p_{\rm s}^\ast$} & \colhead{$p_{\rm s}$} & \colhead{$N_{\rm FFT}$\tablenotemark{b}\, $/ 10^6$} \phm{{spa}}& \colhead{3FGL Name} & \colhead{Pulsar Name} & \colhead{Max.\tablenotemark{a}\; $p_{\rm s}^\ast$} & \colhead{$p_{\rm s}$} & \colhead{$N_{\rm FFT}$\tablenotemark{b}\, $/ 10^6$}
  }
  \startdata
    \textbf{J0002.6$+$6218} & \textbf{J0002$+$6216} & $\mathbf{0.82}$ & $\mathbf{0.96}$ & $\mathbf{38.78}$ \phm{spa}& J1726.6$-$3530c & \nodata & $> 1$ & \nodata & $292.96$ \\[-1.2ex]
    J0212.1$+$5320 & \nodata & $0.68$  & \nodata & $73.31$ \phm{spa}&J1736.0$-$2701 & \nodata & $> 1$ & \nodata & $160.02$ \\[-1.2ex]
    J0223.6$+$6204 & \nodata & $0.65$  & \nodata & $304.66$ \phm{spa}&J1740.5$-$2642 & \nodata & $> 1$ & \nodata & $56.71$ \\[-1.2ex]
    J0225.8$+$6159 & \nodata & $0.96$  & \nodata & $400.20$ \phm{spa}&J1740.5$-$2726 & \nodata & $> 1$ & \nodata & $177.86$ \\[-1.2ex]
    \textbf{J0359.5$+$5413} & \textbf{J0359$+$5414} & $\mathbf{0.78}$ & $\mathbf{0.87}$ & $\mathbf{258.10}$ \phm{spa}& J1740.5$-$2843 & \nodata & $0.87$ & \nodata & $117.70$ \\[-1.2ex]
    J0426.7$+$5437 & \nodata & $0.61$  & \nodata & $45.44$ \phm{spa}&J1742.6$-$3321 & \nodata & $> 1$ & \nodata & $231.76$ \\[-1.2ex]
    J0541.1$+$3553 & \nodata & $> 1$  & \nodata & $776.03$ \phm{spa}&\textbf{J1744.1$-$7619} & \textbf{J1744$-$7619\tablenotemark{c}} & $\mathbf{0.66}$ & $\mathbf{0.72}$ & $\mathbf{275.49}$ \\[-1.2ex]
    \textbf{J0631.6$+$0644} & \textbf{J0631$+$0646} & $\mathbf{0.91}$ & $\mathbf{0.90}$ & $\mathbf{10.11}$ \phm{spa}& J1745.1$-$3011 & \nodata & $0.94$ & \nodata & $165.41$ \\[-1.2ex]
    J0634.1$+$0424 & \nodata & $0.56$  & \nodata & $1140.90$ \phm{spa}&J1745.3$-$2903c & \nodata & $0.40$ & \nodata & $16.65$ \\[-1.2ex]
    J0744.1$-$2523 & \nodata & $0.87$  & \nodata & $26.92$ \phm{spa}&J1746.3$-$2851c & \nodata & $0.48$ & \nodata & $17.16$ \\[-1.2ex]
    J0854.8$-$4503 & \nodata & $0.87$  & \nodata & $579.82$ \phm{spa}&J1747.0$-$2828 & \nodata & $0.60$ & \nodata & $45.69$ \\[-1.2ex]
    J0855.4$-$4818 & \nodata & $> 1$  & \nodata & $4899.56$ \phm{spa}&J1747.7$-$2904 & \nodata & $0.98$ & \nodata & $43.13$ \\[-1.2ex]
    J0901.6$-$4700 & \nodata & $> 1$  & \nodata & $415.84$ \phm{spa}&J1748.3$-$2815c & \nodata & $0.95$ & \nodata & $23.44$ \\[-1.2ex]
    J0907.0$-$4802 & \nodata & $> 1$  & \nodata & $860.15$ \phm{spa}&J1749.2$-$2911 & \nodata & $> 1$ & \nodata & $193.65$ \\[-1.2ex]
    J0933.9$-$6232 & \nodata & $0.90$  & \nodata & $133.98$ \phm{spa}&J1754.0$-$2538 & \nodata & $0.99$ & \nodata & $11.29$ \\[-1.2ex]
    J1026.2$-$5730 & \nodata & $0.97$  & \nodata & $713.27$ \phm{spa}&J1754.0$-$2930 & \nodata & $0.94$ & \nodata & $79.32$ \\[-1.2ex]
    \textbf{J1035.7$-$6720} & \textbf{J1035$-$6720\tablenotemark{c}} & $\mathbf{0.69}$ & $\mathbf{0.93}$ & $\mathbf{353.57}$ \phm{spa}& J1758.8$-$2346 & \nodata & $> 1$ & \nodata & $21.08$ \\[-1.2ex]
    J1037.9$-$5843 & \nodata & $> 1$  & \nodata & $748.28$ \phm{spa}&J1800.8$-$2402 & \nodata & $0.92$ & \nodata & $12.44$ \\[-1.2ex]
    J1039.1$-$5809 & \nodata & $> 1$  & \nodata & $7498.83$ \phm{spa}&J1814.0$-$1757c & \nodata & $> 1$ & \nodata & $33.23$ \\[-1.2ex]
    J1047.3$-$6005 & \nodata & $> 1$  & \nodata & $1607.52$ \phm{spa}&J1814.1$-$1734c & \nodata & $0.98$ & \nodata & $43.69$ \\[-1.2ex]
    J1048.2$-$5928 & \nodata & $> 1$  & \nodata & $651.27$ \phm{spa}&J1823.2$-$1339 & \nodata & $0.71$ & \nodata & $41.60$ \\[-1.2ex]
    \textbf{J1056.7$-$5853} & \textbf{J1057$-$5851} & $\mathbf{0.85}$ & $\mathbf{0.68}$ & $\mathbf{1793.05}$ \phm{spa}& \textbf{J1827.3$-$1446} & \textbf{J1827$-$1446} & $\mathbf{0.95}$ & $\mathbf{1.00}$ & $\mathbf{169.81}$ \\[-1.2ex]
    J1101.9$-$6053 & \nodata & $0.94$  & \nodata & $1676.99$ \phm{spa}&J1831.7$-$0230 & \nodata & $> 1$ & \nodata & $624.00$ \\[-1.2ex]
    \textbf{J1104.9$-$6036} & \textbf{J1105$-$6037} & $\mathbf{0.80}$ & $\mathbf{0.71}$ & $\mathbf{499.71}$ \phm{spa}& J1833.9$-$0711 & \nodata & $> 1$ & \nodata & $27.47$ \\[-1.2ex]
    J1111.9$-$6038 & \nodata & $0.42$  & \nodata & $113.63$ \phm{spa}&J1834.5$-$0841 & \nodata & $> 1$ & \nodata & $208.28$ \\[-1.2ex]
    J1112.0$-$6135 & \nodata & $> 1$  & \nodata & $2341.02$ \phm{spa}&J1839.3$-$0552 & \nodata & $0.78$ & \nodata & $104.52$ \\[-1.2ex]
    J1119.9$-$2204 & \nodata & $0.62$  & \nodata & $13.72$ \phm{spa}&J1840.1$-$0412 & \nodata & $> 1$ & \nodata & $145.45$ \\[-1.2ex]
    J1139.0$-$6244 & \nodata & $> 1$  & \nodata & $42.51$ \phm{spa}&J1843.7$-$0322 & \nodata & $0.84$ & \nodata & $747.26$ \\[-1.2ex]
    \textbf{J1208.4$-$6239} & \textbf{J1208$-$6238\tablenotemark{d}} & $\mathbf{0.77}$ & $\mathbf{0.53}$ & $\mathbf{92.51}$ \phm{spa}& \textbf{J1844.3$-$0344} & \textbf{J1844$-$0346} & $\mathbf{0.95}$ & $\mathbf{0.88}$ & $\mathbf{318.79}$ \\[-1.2ex]
    J1212.2$-$6251 & \nodata & $> 1$  & \nodata & $61.67$ \phm{spa}&J1848.4$-$0141 & \nodata & $0.71$ & \nodata & $727.88$ \\[-1.2ex]
    J1214.0$-$6236 & \nodata & $0.81$  & \nodata & $146.94$ \phm{spa}&J1849.4$-$0057 & \nodata & $0.86$ & \nodata & $176.94$ \\[-1.2ex]
    J1306.4$-$6043 & \nodata & $0.75$  & \nodata & $161.88$ \phm{spa}&J1850.5$-$0024 & \nodata & $> 1$ & \nodata & $458.08$ \\[-1.2ex]
    J1317.6$-$6315 & \nodata & $> 1$  & \nodata & $314.90$ \phm{spa}&J1852.8$+$0158 & \nodata & $0.95$ & \nodata & $379.46$ \\[-1.2ex]
    J1329.8$-$6109 & \nodata & $> 1$  & \nodata & $42.10$ \phm{spa}&J1855.4$+$0454 & \nodata & $> 1$ & \nodata & $95.62$ \\[-1.2ex]
    J1345.1$-$6224 & \nodata & $> 1$  & \nodata & $182.65$ \phm{spa}&J1857.2$+$0059 & \nodata & $> 1$ & \nodata & $220.81$ \\[-1.2ex]
    \textbf{J1350.4$-$6224} & \textbf{J1350$-$6225} & $\mathbf{> 1}$ & $\mathbf{1.00}$ & $\mathbf{100.80}$ \phm{spa}& J1857.8$+$0129c & \nodata & $> 1$ & \nodata & $109.05$ \\[-1.2ex]
    J1358.5$-$6025 & \nodata & $0.87$  & \nodata & $199.81$ \phm{spa}&J1857.9$+$0210 & \nodata & $0.89$ & \nodata & $459.00$ \\[-1.2ex]
    J1405.4$-$6119 & \nodata & $0.58$  & \nodata & $198.20$ \phm{spa}&J1857.9$+$0355 & \nodata & $> 1$ & \nodata & $270.39$ \\[-1.2ex]
    J1503.5$-$5801 & \nodata & $> 1$  & \nodata & $459.50$ \phm{spa}&J1859.6$+$0102 & \nodata & $> 1$ & \nodata & $120.74$ \\[-1.2ex]
    \textbf{J1528.3$-$5836} & \textbf{J1528$-$5838} & $\mathbf{> 1}$ & $\mathbf{0.98}$ & $\mathbf{21.22}$ \phm{spa}& J1900.8$+$0337 & \nodata & $> 1$ & \nodata & $354.12$ \\[-1.2ex]
    J1539.2$-$3324 & \nodata & $> 1$  & \nodata & $7.42$ \phm{spa}&J1901.1$+$0728 & \nodata & $> 1$ & \nodata & $278.80$ \\[-1.2ex]
    J1549.1$-$5347c & \nodata & $0.84$  & \nodata & $663.95$ \phm{spa}&\textbf{J1906.6$+$0720} & \textbf{J1906$+$0722\tablenotemark{e}} & $\mathbf{0.58}$ & $\mathbf{0.77}$ & $\mathbf{206.02}$ \\[-1.2ex]
    J1552.8$-$5330 & \nodata & $> 1$  & \nodata & $930.32$ \phm{spa}&J1910.9$+$0906 & \nodata & $0.38$ & \nodata & $47.32$ \\[-1.2ex]
    J1620.0$-$5101 & \nodata & $> 1$  & \nodata & $384.50$ \phm{spa}&J1919.9$+$1407 & \nodata & $> 1$ & \nodata & $383.20$ \\[-1.2ex]
    \textbf{J1622.9$-$5004} & \textbf{J1623$-$5005} & $\mathbf{0.76}$ & $\mathbf{0.74}$ & $\mathbf{78.62}$ \phm{spa}& J1925.4$+$1727 & \nodata & $> 1$ & \nodata & $1704.54$ \\[-1.2ex]
    \textbf{J1624.2$-$4041} & \textbf{J1624$-$4041} & $\mathbf{0.74}$ & $\mathbf{0.85}$ & $\mathbf{192.42}$ \phm{spa}& J1928.9$+$1739 & \nodata & $> 1$ & \nodata & $2680.58$ \\[-1.2ex]
    J1625.1$-$0021 & \nodata & $0.69$  & \nodata & $154.54$ \phm{spa}&J2004.4$+$3338 & \nodata & $0.86$ & \nodata & $84.87$ \\[-1.2ex]
    J1626.2$-$2428c & \nodata & $> 1$  & \nodata & $10.64$ \phm{spa}&\textbf{J2017.9$+$3627} & \textbf{J2017$+$3625} & $\mathbf{0.53}$ & $\mathbf{0.64}$ & $\mathbf{178.21}$ \\[-1.2ex]
    J1636.2$-$4709c & \nodata & $> 1$  & \nodata & $405.59$ \phm{spa}&J2023.5$+$4126 & \nodata & $> 1$ & \nodata & $776.94$ \\[-1.2ex]
    J1636.2$-$4734 & \nodata & $0.66$  & \nodata & $321.30$ \phm{spa}&J2032.5$+$3921 & \nodata & $> 1$ & \nodata & $541.70$ \\[-1.2ex]
    J1638.6$-$4654 & \nodata & $0.89$  & \nodata & $102.88$ \phm{spa}&J2034.6$+$4302 & \nodata & $> 1$ & \nodata & $1601.98$ \\[-1.2ex]
    J1639.4$-$5146 & \nodata & $0.77$  & \nodata & $9.24$ \phm{spa}&J2035.0$+$3634 & \nodata & $> 1$ & \nodata & $26.52$ \\[-1.2ex]
    J1641.1$-$4619c & \nodata & $> 1$  & \nodata & $14.19$ \phm{spa}&J2038.4$+$4212 & \nodata & $> 1$ & \nodata & $845.17$ \\[-1.2ex]
    J1650.0$-$4438c & \nodata & $> 1$  & \nodata & $161.85$ \phm{spa}&J2039.4$+$4111 & \nodata & $> 1$ & \nodata & $525.98$ \\[-1.2ex]
    \textbf{J1650.3$-$4600} & \textbf{J1650$-$4601} & $\mathbf{0.74}$ & $\mathbf{0.78}$ & $\mathbf{139.09}$ \phm{spa}& J2039.6$-$5618 & \nodata & $0.74$ & \nodata & $91.84$ \\[-1.2ex]
    J1652.8$-$4351 & \nodata & $> 1$  & \nodata & $741.14$ \phm{spa}&J2041.1$+$4736 & \nodata & $0.71$ & \nodata & $170.62$ \\[-1.2ex]
    J1702.8$-$5656 & \nodata & $0.52$  & \nodata & $96.18$ \phm{spa}&J2042.4$+$4209 & \nodata & $> 1$ & \nodata & $5359.83$ \\[-1.2ex]
    J1714.5$-$3832 & \nodata & $0.48$  & \nodata & $67.93$ \phm{spa}&J2112.5$-$3044 & \nodata & $0.69$ & \nodata & $22.79$ \\[-1.2ex]
    J1718.0$-$3726 & \nodata & $0.96$  & \nodata & $2.56$ \phm{spa}&J2323.4$+$5849 & \nodata & $0.57$ & \nodata & $63.63$ \\
  \enddata
  \tablecomments{Sources in which pulsars were discovered by the \EAH{} survey are shown in bold. For some of these pulsars, the measured pulsed fraction is well below our estimated upper limit. This can be due to the pulsar having a narrower-than-average pulse profile, a very low spin frequency (at which the sky grid, constructed for the highest frequency in the search band greatly overcovers the search region), or simple ``luck'' in that the signal lay close to one of our search points and had a lower-than-average mismatch. The $95\%$ detection probability requirement therefore results in conservative limits.}
  \tablenotetext{a}{Estimated values for the pulsed fraction above which we expect to detect a signal from each source with 95\% probability.}
  \tablenotetext{b}{Number of FFTs required to search the entire parameter space for each source.}
  \tablenotetext{c}{These pulsars have timing properties warranting further individual
    investigation, and will be presented in later works.}
  \tablenotetext{d}{The discovery and analysis of PSR~J1208$-$6238 was presented in \citet{Clark2016+J1208}.}
  \tablenotetext{e}{The discovery and analysis of PSR~J1906$+$0722 was presented in \citet{Clark2015+J1906}.}
\end{deluxetable}
\null
\clearpage
We have also not considered the fact that our sensitivity is not exactly constant throughout the
parameter space. Rather, the grid of sky locations slightly overcovers the region at low frequencies and,
due to the lower number of sky points required at low frequencies, work units searching the lowest
frequency band often search at a smaller number of trials. The result is that our
survey is sometimes slightly more sensitive at low spin frequencies. The results of Table
\ref{t:minimum_pulsed_fraction} have been averaged over all spin frequencies, and assume that the
entire parameter space is equally well-covered by search points.

In the following sections we highlight and discuss the implications of our measured pulsed fraction
upper limits for three sources with well-known counterparts from observations
at other wavelengths, from which no pulsations were detected by our survey.
\subsection{Pulsed fraction upper limit for W49B}
\textit{Chandra} observations of the \ac{SNR} W49B (3FGL~J1910.9$+$0906), believed to be the remnant of a
jet-driven, core-collapse supernova, place strong upper limits on the presence of a neutron star
\citep{Lopez2013+ChandraW49B}.

Gamma-rays from W49B have been detected at energies far higher than observed from a typical
gamma-ray pulsar \citep{Abdo2010+W49B,HESS2016+W49B}, indicating that
the majority of the high-energy flux from W49B comes from the SNR itself. Any pulsed
emission from a gamma-ray pulsar would therefore likely represent only a small fraction of the
photon flux. Our results place a 95\% pulsed fraction limit of $38\%$ of the weighted photon
flux from this source.

Placing W49B at the distance of $10$ kpc obtained by \citet{Zhu2014+W49B} suggests a gamma-ray
luminosity of $\sim 2\times10^{36}$~erg~s$^{-1}$. In order to provide a significant fraction of this
emission, any gamma-ray pulsar would have to have a very large spin-down power, since the efficiency
of converting spin-down power into gamma-ray luminosity tends to be lower for energetic pulsars
\citep{2PC+2013}. The estimated age of the \ac{SNR} is in the range $1000$--$4000$~yr. Together, these
observations suggest that any potential gamma-ray pulsar would be very young and extremely
energetic, and would likely exhibit large timing noise and glitches as a result. This would
seriously reduce the sensitivity of our blind search of this target, making our upper limit estimate
for the pulsed flux unreliable for this source.

To check for signals with large timing noise, we manually followed up semicoherent candidates
from this source using refinement steps with increasing lag-window lengths, but none revealed a
significant pulsed signal.

\subsection{Pulsed fraction upper limit for  Cassiopeia A}
The \ac{SNR} Cassiopeia A (Cas A) contains a point-like, central X-ray source, most likely a neutron
star \citep{Ho2009+CasANS}, from which no pulsations have been detected in X-rays, gamma rays or
radio observations. The gamma-ray spectrum for this source is also unlike that of any pulsar, again
suggesting that any pulsed emission would likely only contribute a fraction of the total observed
flux. The position of this central compact object is within our search region for 3FGL~J2323.4$+$5849, a source for which our survey sets a pulsed fraction upper limit of $p_s^{\ast} =
57\%$. The photon flux above $100$~MeV from this source was $3.1(2) \times
10^{-8}$~photons~cm$^{-2}$~s$^{-1}$, making our 95\% upper limit more than an order of magnitude
lower than the 5$\sigma$ limit for pulsed flux reported in \citet{Abdo2010+CasA}. A dedicated search
for pulsations at the known position of the Cas A neutron star, excluding photons above typical
pulsar emission energies, could likely bring this limit down further.

However, since Cas A is known to be a young SNR (the supernova occurred
around A.D. 1680), if the NS is indeed a pulsar, it will be very energetic and likely have a highly
unstable spin, making detection in a blind search extremely challenging even if the pulsed fraction
is far higher than our stated upper limit. Indeed, the SNR is young enough that the
pulsar's spin-down could even be outside our search range \citep{Abadie2010+CasA}. Again, we
followed up semicoherent candidates from this source, without success.

\subsection{Pulsed fraction upper limit for the Galactic Center}
As a result of intense and difficult-to-model interstellar emission, the area around the Galactic
Center (GC) is one of the most complicated, and hence poorly understood regions of gamma-ray
emission. Both the 3FGL and the recent First \Fermi-LAT Inner Galaxy Point Source Catalog
\citep{Ajello2016+1FIG} identify several bright point sources within a few degrees of the GC; though
some of these could be due to misattributed interstellar emission. Nevertheless, a substantial
contribution to the flux from the GC region is expected to come from other astrophysical sources,
such as young pulsars or MSPs \citep[e.g.][submitted, and references
  therein]{Bartels2015+GCMSPs,OLeary2016+GCPSRs}, or possibly even annihilating dark matter
particles \citep[][and references therein]{Hooper2013+GCDM}. The detection of a gamma-ray pulsar
near the GC would have important implications for these two competing interpretations of the GC GeV
flux.

The bright fore-/background from the interstellar medium makes blindly searching for pulsars near
the GC particularly difficult. In order for one single pulsar to be detectable above
this background, it must be extremely bright, especially if it lies at a similar distance as the GC,
$\sim 8$~kpc. As an example, the bright source 3FGL~J1745.3$-$2903c searched during this survey has
$p_{\rm s}^{\ast} =  40\%$, corresponding to a pulsed photon flux above 1GeV of $\sim
10^{-8}$~photons~cm$^{-2}$~s$^{-1}$. This flux is similar to the photon flux that the Crab pulsar
would produce if it was at the same distance as the GC. Again, such highly luminous pulsars also
exhibit the most timing noise and glitches, further adding to the difficulty of detecting their
pulsations above the bright background flux.

\section{Timing Analysis}\label{s:timing}
We performed timing analyses for each new pulsar to precisely determine their sky positions
and rotational parameters, again denoted by the vector $\boldsymbol{u}$. The analysis follows the
procedure described in \citet{Clark2015+J1906}, as an extension of the method described by
\citet{Ray2011}.

For the purpose of these follow-up timing analyses, we produced extended LAT data sets until
2015~September~9 for each of the pulsars newly reported in this work. These updated data sets were
produced using the \texttt{P8R2\_SOURCE\_V6} IRFs, \texttt{gll\_iem\_v06.fits} Galactic diffuse
emission template \citep{Acero2016+Diffuse}, and \texttt{iso\_P8R2\_SOURCE\_V6\_v06.txt} isotropic
diffuse background
template.\footnote{\url{http://fermi.gsfc.nasa.gov/ssc/data/access/lat/BackgroundModels.html}} The
extended data sets had a lower zenith angle cutoff of $90\arcdeg$. The pulsar's position was fixed
at its initial timing position. Photons from within a larger $15\arcdeg$ radius were included in the
likelihood fitting, which was performed using \texttt{gtlike}. Photon weights were then calculated
for all photons from within $5\arcdeg$ of the pulsar using \texttt{gtsrcprob}. Further details of
the preparation of these data will be given in Paper~II, including the spectral properties of each
newly detected pulsar.

To reduce the number of photons included in the timing analysis for computational efficiency, we
applied a photon weight cutoff with the minimum photon weight chosen such that no more than $1\%$ of
the maximum coherent Fourier power was lost (again assuming that the maximum coherent S/N is
proportional to $\sum_{j=1}^N w_j^2$).

Starting from the spin and positional parameters of the pulsar reported by the refinement stage, we
phase-folded the photon data to obtain a weighted pulse profile. We also phase-folded at half, and one-third of the measured frequency to ensure that the original signal was not a higher harmonic of the
fundamental spin frequency. In two pulsars, J1350$-$6225 and J1624$-$4041, this revealed
sharply double-peaked profiles at half of the original candidate frequency, and greatly increased
their measured $H$-test values, indicative of having identified the true spin frequency.

\begin{figure*}
  \centering
	\includegraphics[width=0.9\textwidth]{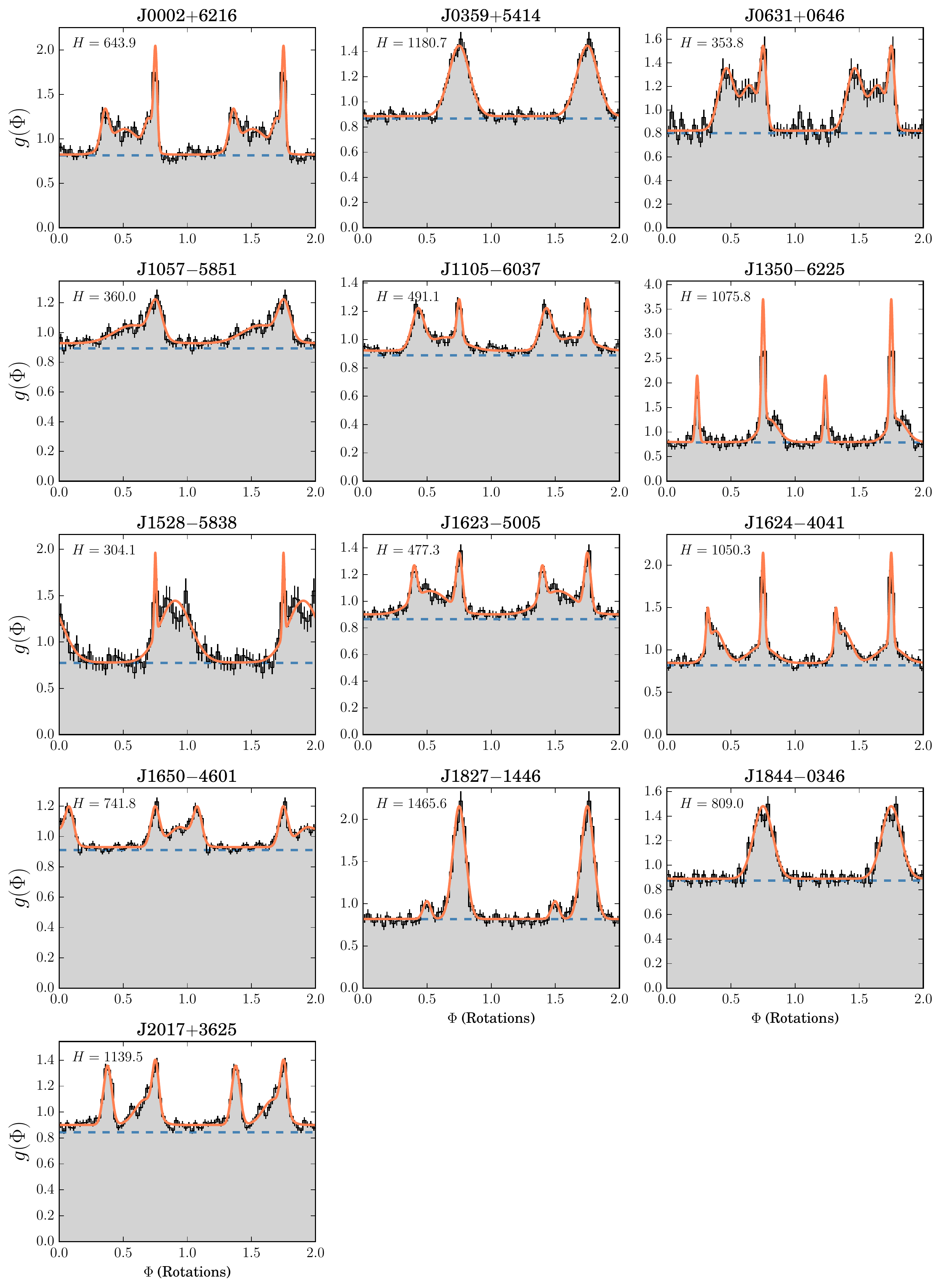}
	\caption{Weighted pulse profiles of all pulsars reported in this work. The template pulse
    profiles used for timing analyses are shown by orange curves. The background fraction is
    illustrated by the dashed blue line in each plot.}
	\label{f:all_pulse_profiles}
\end{figure*}

From the phase-folded data, we constructed a template pulse profile, $\hat{g}_s\left(\Phi\right)$,
consisting of a combination of symmetrical wrapped Gaussian peaks (as defined in \citet{2PC+2013}),
which were fit by maximizing the likelihood,
\begin{equation}
  \mathcal{L}\left(\hat{g}_s, \boldsymbol{u}\right) = \prod_{j=1}^N \left[ w_j\,
    \hat{g}_s\left(\Phi(t_j,\boldsymbol{u})\right) + (1 - w_j)\right]\,.
  \label{e:likelihood}
\end{equation}
The number of peaks in the template profile was chosen by the template that minimizes the Bayesian
Information Criterion \citep[BIC,][]{Schwarz1978+BIC},
\begin{equation}
  BIC = -2 \log\left(\mathcal{L}\left(\hat{g}_s, \boldsymbol{u}\right)\right) +
  k\log\left(\sum_{j=1}^N w_j\right)\,,
  \label{e:BIC}
\end{equation}
where $k$ is the number of free parameters in the model. Because each Gaussian peak consists of
three parameters (central phase, width, and amplitude), when fitting the template pulse profile, $k =
3 N_{\rm peaks}$. Due to the presence of the second term in Equation~(\ref{e:BIC}), a new component
was only added to the template profile if its presence significantly increased the likelihood. It
therefore acts as a penalty factor, discriminating against a template profile featuring many
``spiky'' components, unless this is warranted by the data. The parameters of the template pulse
profiles used to time each pulsar are given in Table~\ref{t:pulse_profiles_table}, and the profiles
themselves are shown in Figure~\ref{f:all_pulse_profiles}.

After obtaining the template pulse profile, we varied the positional and spin parameters and
explored the resulting multi-dimensional likelihood surface to find the most likely parameter
values. To explore the likelihood surface, we used the \ac{AIMC} method described by
\citet{Goodman2010+AIMC}, in which many Monte~Carlo chains are run in parallel, with proposal jumps
for each chain depending on the locations of the other chains. We used the scheme described by
\citet{Foreman-Mackey2013+emcee} to efficiently parallelize the likelihood computations amongst
several CPU cores.

\begin{deluxetable*}{rrrrrrrrrrrr}
  \tablecaption{Template Pulse Profile Parameters \label{t:pulse_profiles_table}}
  \tabletypesize{\footnotesize}
  \tablecolumns{12}
  \tablehead{
    \colhead{Pulsar} & \colhead{$a_1$} & \colhead{$\sigma_1$} & \colhead{$\Delta \mu_2$} & \colhead{$a_2$} & \colhead{$\sigma_2$} & \colhead{$\Delta \mu_3$} & \colhead{$a_3$} & \colhead{$\sigma_3$} & \colhead{$\Delta \mu_4$} & \colhead{$a_4$} & \colhead{$\sigma_4$}}
  \startdata
    J0002$+$6216 & $0.32$ & $0.088$ \phs & $-2.476$ & $0.25$ & $0.186$ \phs & $-0.319$ & $0.26$ & $0.217$ \phs & $-1.531$ & $0.68$ & $0.697$\\[-0.75ex]
    J0359$+$5414 & $1.36$ & $0.511$ \phs & \nodata & \nodata & \nodata \phs & \nodata & \nodata & \nodata \phs & \nodata & \nodata & \nodata\\[-0.75ex]
    J0631$+$0646 & $0.29$ & $0.145$ \phs & $-1.821$ & $0.62$ & $0.374$ \phs & $-0.663$ & $0.50$ & $0.408$ \phs & \nodata & \nodata & \nodata\\[-0.75ex]
    J1057$-$5851 & $0.43$ & $0.301$ \phs & $-1.143$ & $0.65$ & $0.907$ \phs & \nodata & \nodata & \nodata \phs & \nodata & \nodata & \nodata\\[-0.75ex]
    J1105$-$6037 & $0.19$ & $0.116$ \phs & $-2.026$ & $0.46$ & $0.294$ \phs & $-0.632$ & $0.47$ & $0.866$ \phs & \nodata & \nodata & \nodata\\[-0.75ex]
    J1350$-$6225 & $0.57$ & $0.077$ \phs & $3.053$ & $0.31$ & $0.079$ \phs & $0.278$ & $0.62$ & $0.453$ \phs & \nodata & \nodata & \nodata\\[-0.75ex]
    J1528$-$5838 & $0.13$ & $0.053$ \phs & $0.972$ & $1.42$ & $0.762$ \phs & \nodata & \nodata & \nodata \phs & \nodata & \nodata & \nodata\\[-0.75ex]
    J1623$-$5005 & $0.30$ & $0.151$ \phs & $-2.215$ & $0.17$ & $0.141$ \phs & $-1.438$ & $0.70$ & $0.855$ \phs & \nodata & \nodata & \nodata\\[-0.75ex]
    J1624$-$4041 & $0.31$ & $0.080$ \phs & $-2.715$ & $0.11$ & $0.075$ \phs & $-2.328$ & $0.50$ & $0.383$ \phs & $-0.072$ & $0.42$ & $0.604$\\[-0.75ex]
    J1650$-$4601 & $0.38$ & $0.210$ \phs & $2.079$ & $0.33$ & $0.211$ \phs & $1.227$ & $0.52$ & $0.571$ \phs & \nodata & \nodata & \nodata\\[-0.75ex]
    J1827$-$1446 & $1.44$ & $0.311$ \phs & $-1.581$ & $0.13$ & $0.182$ \phs & \nodata & \nodata & \nodata \phs & \nodata & \nodata & \nodata\\[-0.75ex]
    J1844$-$0346 & $1.38$ & $0.467$ \phs & \nodata & \nodata & \nodata \phs & \nodata & \nodata & \nodata \phs & \nodata & \nodata & \nodata\\[-0.75ex]
    J2017$+$3625 & $0.37$ & $0.200$ \phs & $2.356$ & $0.25$ & $0.160$ \phs & $1.843$ & $0.39$ &
    $0.494$ \phs & \nodata & \nodata & \nodata
  \enddata
  \tablecomments{Columns $2$--$12$ give the amplitude ($a_i$), offset in radians from the tallest component $\left(\Delta \mu_i\right)$, and width parameter ($\sigma_i$) for each wrapped Gaussian component in the template pulse profile.}
\end{deluxetable*}
With the new parameter values, we re-folded the photon data to obtain a new template pulse profile as
above. Additional parameters could then be added to the timing model, and the procedure was
repeated. For each pulsar, we started from the simplified timing model (i.e. up to $\dot{f}$), added
higher frequency derivative terms, found the most likely parameters, and updated the template pulse
profile until the BIC of Equation (\ref{e:BIC}) (with $k$ now as the number of parameters in the
timing model) stopped decreasing. With the number of timing parameters selected in this manner, we
performed a final longer Monte~Carlo run, using a large number of chains, to obtain precise
estimates of the mean value and uncertainty of each parameter.

With over $100,000$ photons whose individual barycentric arrival times must be computed each time,
each likelihood evaluation is relatively computationally expensive. Hence, efficient convergence of
the Monte Carlo step is crucial to allow us to perform the timing analysis in a reasonable amount of
time. To avoid the possibility of chains getting stuck in low-likelihood regions, we start all of
the chains in a tight ball near our current most-likely point, as advocated by
\citet{Foreman-Mackey2013+emcee}. Using a lengthy burn-in period, we allow these chains to spread
out throughout the most likely regions of our parameter space. While this initialization can, in
principle, lead to the Monte Carlo sampling only reaching a local likelihood maximum rather than
exploring the full parameter space to find a global maximum, visual inspection of the phase-folded
photon data can typically reveal any significant residuals in the timing solution requiring further
fitting.

The results of these analyses are summarized in Table \ref{t:params1}. The physical properties of
each of the new pulsars, as derived from their spin frequency and spin-down rate are given in Table
\ref{t:params2}.

These timing solutions allow for sensitive follow-up searches, the identification of candidate
multiwavelength counterparts, and phase-resolved spectroscopy of the on- and off-pulse
photons. Dedicated radio observations of the newly discovered pulsars were also performed, which
used the gamma-ray timing solution to fold the data. For those pulsars that were subsequently
detected in radio observations, the phase alignment between the radio and gamma-ray pulses provide
constraints on the pulsars' emission geometry and inclination, allowing for the comparison of
different gamma-ray emission models \citep[e.g.][]{Johnson2014+LCModelling}. These analyses and
their results will be described in Paper~II.

\subsection{Spin-down vs. Timing Noise}
\label{s:timing_noise}
The long-term spin-down behavior of a pulsar can be characterized by the braking index
\citep[e.g.,][]{Lyne2015+Crab}, $n$, where,
\begin{equation}
  \dot{f} \propto -f^{n}\,,
\end{equation}
\begin{equation}
  n = \frac{f \, \ddot{f}}{\dot{f}^2}\,.
  \label{e:braking_index}
\end{equation}
The exact value of the braking index depends on the physical mechanism causing the pulsar to spin
down; a pulsar whose braking is entirely due to magnetic dipole radiation will have $n = 3$, whereas
one whose spin-down power is entirely due to the radiation of gravitational waves will have
$n=5$ or $n=7$ \citep{Abadie2010+CasA}.

The vast majority of pulsars, however, also exhibit red-spectrum ``timing noise'',
manifesting as low-frequency quasi-periodic variations in the arrival times of pulses
\citep[e.g.,][]{Hobbs2010+TimingNoise,Kerr2015+FermiTiming}. The amplitude of this timing noise
appears to correlate with the spin-down energy, $\dot{E}$, which is typically higher for gamma-ray
pulsars than the rest of the pulsar population. For all but the youngest pulsars or those with
the highest magnetic fields, this timing noise obscures the true long time-scale braking behavior.

\begin{figure*}
  \centering
	\includegraphics[width=0.95\textwidth]{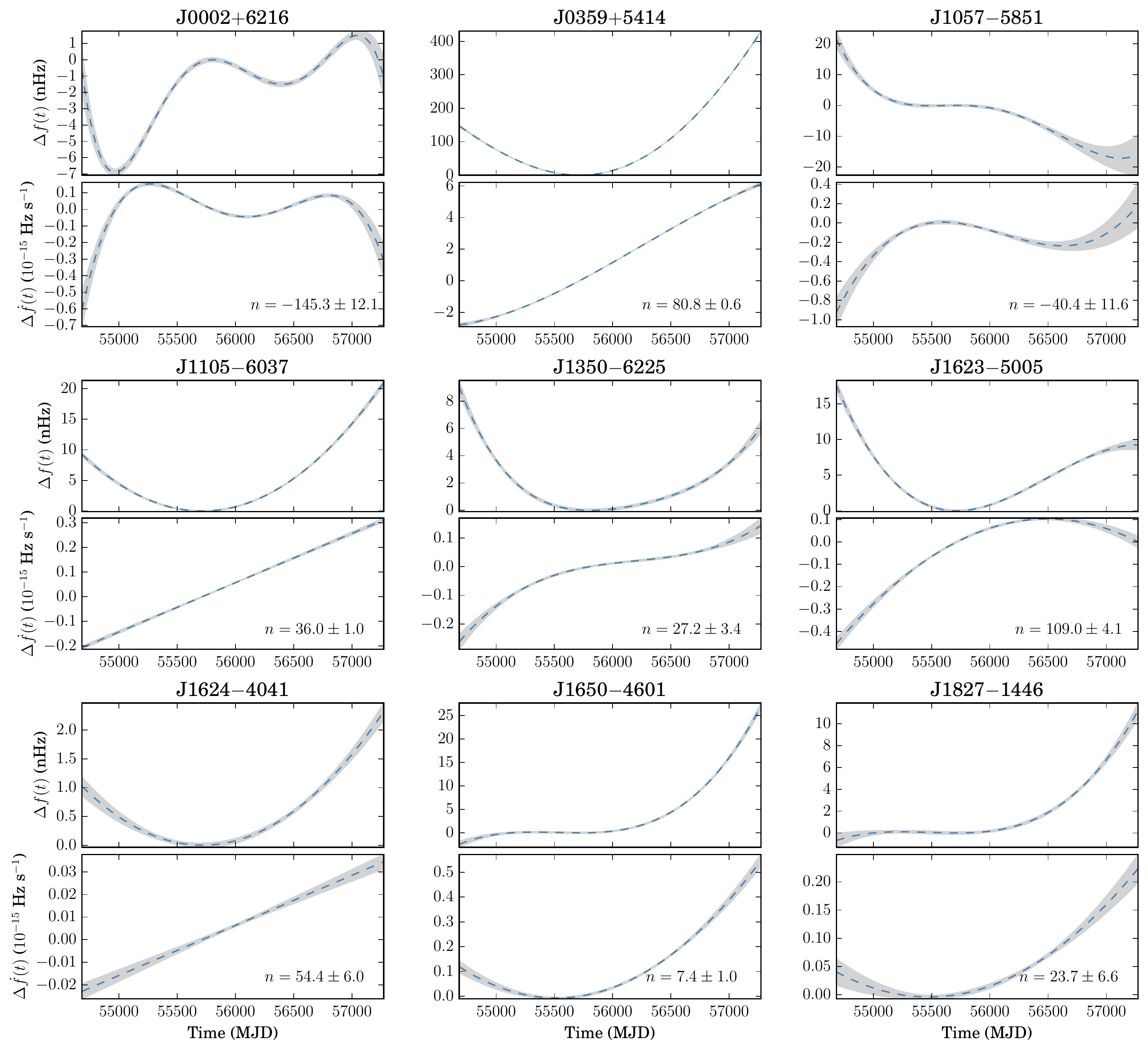}
	\caption{Evolution of each pulsar's rotational frequency during the \Fermi-LAT observation
    time. For each pulsar, the top and bottom panels show the deviations from a constant spin-down
    model of the frequency, $\Delta f(t)$, and first frequency derivative, $\Delta \dot{f} (t)$,
    respectively. The shaded areas show 1$\sigma$ uncertainty regions. These deviations are most
    likely due to the pulsars' intrinsic timing noise, as suggested by their unphysical braking
    indices ($n$, described in Section \ref{s:timing_noise} and shown in the lower panel for each
    pulsar).}
	\label{f:all_freq_evos}
\end{figure*}

In all pulsars measured here, $n$ deviates significantly from any of these integer values (except
for PSR~J1650$-$4601, where the index is low, but poorly constrained), suggesting that the measured
values of $\ddot{f}$ are due to short time-scale timing noise. For pulsars with measurable frequency
derivative terms beyond the first derivative, the evolution of the spin frequency and spin-down rate
is shown in Figure \ref{f:all_freq_evos}.
\onecolumngrid
\vspace{2cm}
\begin{deluxetable*}{llllDD}
  \tablecaption{Pulsar timing parameters \label{t:params1}}
  \tablecolumns{6}
  \tablewidth{0pt}
  \tablehead{
    \colhead{Pulsar} & \colhead{$t_{\rm ref}$ (MJD)} & \colhead{R.A.} & \colhead{Decl.} & \twocolhead{$f$ (Hz)} & \twocolhead{$\dot{f}$ ($10^{-12}$ Hz s$^{-1}$)}
  }
  \decimals
  \startdata
J0002$+$6216 & $55806$ & $00^{\rm h}\,02^{\rm m}\,58^{\rm s}.17(2)$ &$+62\,\arcdeg16\,\arcmin09\farcs4(1)$ &$8.6682478274(1)$ &$-0.448354(5)$ \\
J0359$+$5414 & $55716$ & $03^{\rm h}\,59^{\rm m}\,26^{\rm s}.01(2)$ &$+54\,\arcdeg14\,\arcmin55\farcs7(3)$ &$12.5901403227(2)$ &$-2.65247(1)$ \\
J0631$+$0646 & $55806$ & $06^{\rm h}\,31^{\rm m}\,52^{\rm s}.38(2)$ &$+06\,\arcdeg46\,\arcmin14\farcs2(7)$ &$9.01071834910(6)$ &$-0.293694(2)$ \\
J1057$-$5851 & $55716$ & $10^{\rm h}\,57^{\rm m}\,09^{\rm s}.5(2)$ &$-58\,\arcdeg51\,\arcmin07(2)\arcsec$ &$1.6119541713(3)$ &$-0.26135(1)$ \\
J1105$-$6037 & $55716$ & $11^{\rm h}\,05^{\rm m}\,00^{\rm s}.48(4)$ &$-60\,\arcdeg37\,\arcmin16\farcs3(3)$ &$5.12982912390(8)$ &$-0.574649(2)$ \\
J1350$-$6225 & $55806$ & $13^{\rm h}\,50^{\rm m}\,44^{\rm s}.45(1)$ &$-62\,\arcdeg25\,\arcmin43\farcs8(1)$ &$7.23810134280(6)$ &$-0.465408(2)$ \\
J1528$-$5838 & $55806$ & $15^{\rm h}\,28^{\rm m}\,24^{\rm s}.3(1)$ &$-58\,\arcdeg38\,\arcmin01(1)\arcsec$ &$2.81146362521(6)$ &$-0.195700(1)$ \\
J1623$-$5005 & $55716$ & $16^{\rm h}\,23^{\rm m}\,04^{\rm s}.11(1)$ &$-50\,\arcdeg05\,\arcmin15\farcs1(2)$ &$11.7547287226(1)$ &$-0.574965(3)$ \\
J1624$-$4041 & $55716$ & $16^{\rm h}\,24^{\rm m}\,09^{\rm s}.927(9)$ &$-40\,\arcdeg41\,\arcmin29\farcs7(3)$ &$5.95730476591(3)$ &$-0.1676839(9)$ \\
J1650$-$4601 & $55716$ & $16^{\rm h}\,50^{\rm m}\,18^{\rm s}.62(2)$ &$-46\,\arcdeg01\,\arcmin18\farcs6(4)$ &$7.8664037135(1)$ &$-0.937157(3)$ \\
J1827$-$1446 & $55716$ & $18^{\rm h}\,27^{\rm m}\,24^{\rm s}.60(5)$ &$-14\,\arcdeg46\,\arcmin28(4)\arcsec$ &$2.0032588600(1)$ &$-0.181932(3)$ \\
J1844$-$0346 & $55716$ & $18^{\rm h}\,44^{\rm m}\,32^{\rm s}.89(2)$ &$-03\,\arcdeg46\,\arcmin30\farcs6(9)$ &$8.8609552273(8)$ &$-12.14675(5)$ \\
J2017$+$3625 & $55716$ & $20^{\rm h}\,17^{\rm m}\,55^{\rm s}.84(1)$ &$+36\,\arcdeg25\,\arcmin07\farcs9(2)$ &$5.99703102436(3)$ &$-0.0489063(8)$ \\
  \enddata
 \tablecomments{Reported values of timing parameters are the mean values obtained from the
  Monte-Carlo analysis described in Section \ref{s:timing} at the reference epoch, $t_{\rm ref}$,
  with $1\sigma$ uncertainties in the final digits given in brackets. Reference epochs are in
  Barycentric Dynamical Time (TDB). Two different observation spans were used during this survey and the reference epochs were chosen to lie at the middle of the observation, hence the two distinct values shown in column 2.}
\end{deluxetable*}
\twocolumngrid
\begin{deluxetable*}{lDDDDDDD}
  \tablecaption{Derived pulsar properties \label{t:params2}}
  \tablecolumns{8}
  \tablewidth{0pt}
  \tablehead{
    \colhead{Pulsar} & \twocolhead{$l$ ($\arcdeg$)} & \twocolhead{$b$ ($\arcdeg$)} & \twocolhead{$P$ (ms)} & \twocolhead{$\dot{P}$ ($10^{-15}$ s s$^{-1}$)} & \twocolhead{$\tau_{\rm c}$ (kyr)} & \twocolhead{$\dot{E}$ ($10^{33}$ erg s$^{-1}$)} & \twocolhead{$B_{\rm S}$ ($10^{12}$ G)}
  }
  \decimals
  \startdata
J0002$+$6216 & $117.33$ & $-0.07$ & $115.363568268(2)$ &$5.96703(7)$ &$306$ &$153$ &$0.8$ \\
J0359$+$5414 & $148.23$ & $+0.88$ & $79.427232292(1)$ &$16.73359(7)$ &$75$ &$1318$ &$1.2$ \\
J0631$+$0646 & $204.68$ & $-1.24$ & $110.9789432160(7)$ &$3.61723(2)$ &$486$ &$104$ &$0.6$ \\
J1057$-$5851 & $288.61$ & $+0.80$ & $620.3650313(1)$ &$100.583(5)$ &$98$ &$17$ &$8.0$ \\
J1105$-$6037 & $290.24$ & $-0.40$ & $194.938267113(3)$ &$21.83720(6)$ &$141$ &$116$ &$2.1$ \\
J1350$-$6225 & $309.73$ & $-0.34$ & $138.157778213(1)$ &$8.88352(4)$ &$246$ &$133$ &$1.1$ \\
J1528$-$5838 & $322.17$ & $-1.75$ & $355.686622097(8)$ &$24.7586(2)$ &$228$ &$22$ &$3.0$ \\
J1623$-$5005 & $333.72$ & $-0.31$ & $85.0721461635(8)$ &$4.16118(2)$ &$324$ &$267$ &$0.6$ \\
J1624$-$4041 & $340.56$ & $+6.15$ & $167.861145148(1)$ &$4.72489(2)$ &$563$ &$39$ &$0.9$ \\
J1650$-$4601 & $339.78$ & $-0.95$ & $127.122893310(2)$ &$15.14468(6)$ &$133$ &$291$ &$1.4$ \\
J1827$-$1446 & $17.08$ & $-1.50$ & $499.18661037(3)$ &$45.3351(9)$ &$174$ &$14$ &$4.8$ \\
J1844$-$0346 & $28.79$ & $-0.19$ & $112.85464991(1)$ &$154.7031(6)$ &$12$ &$4249$ &$4.2$ \\
J2017$+$3625 & $74.51$ & $+0.39$ & $166.7491790419(8)$ &$1.35985(2)$ &$1943$ &$12$ &$0.5$ \\
  \enddata
  \tablecomments{Columns 2 and 3 give the pulsars' Galactic longitudes ($l$) and latitudes ($b$) respectively. Columns 4 and 5 give the derived spin period $\left(P = 1/f\right)$ and period derivative $\left(\dot{P} = -\dot{f}/f^2\right)$. Characteristic ages, $\tau_{\rm c}$, spin-down luminosities, $\dot{E}$, and surface magnetic field strengths, $B_{\rm S}$ are calculated as described in \citet{2PC+2013}.}
\end{deluxetable*}

\clearpage

\subsection{Timing Large Glitches}\label{s:J1844}
In addition to slowly varying timing noise behavior, young pulsars also occasionally exhibit large,
sudden changes in their spin frequency, known as ``glitches'' \citep[e.g.][]{Yu2013+Glitches}. Pulsars with
large glitches are particularly difficult to detect in blind searches, which require long
intervals containing a stable signal to accumulate sufficient S/N.

Not only are pulsars with large glitches harder to detect in a blind search, they are also
significantly more complicated to time. A glitch occurring at time $t_{\rm g}$, with increments in
the spin frequency ($\Delta f_{\rm g}$), the first two frequency derivatives ($\Delta \dot{f}_{\rm
  g}$, $\Delta \ddot{f}_{\rm g}$) and an exponentially decaying frequency increment ($\Delta f_{\rm
  D,g}$, decay timescale $\tau_{\rm D,g}$) causes a phase offset at time $t > t_{\rm g}$ of
\begin{multline}
  \Delta \Phi_{\rm g}\left(t\right) = 2\pi\,\left[\Delta f_{\rm g}(t - t_{\rm g}) + \frac{\Delta
      \dot{f}_{\rm g}}{2}(t - t_{\rm g})^2 + \right. \\ \left.\vphantom{ \frac{\Delta \dot{f}_{\rm
          g}}{2}} \frac{\Delta \ddot{f}_{\rm g}}{6} (t - t_{\rm g})^3 + \Delta f_{\rm D, g}
    \tau_{\rm D,g} \left( 1 - {\rm e}^{-(t - t_{\rm g})/\tau_{\rm D,g}}\right)\right]\,.
\end{multline}
If the parameter increments associated with the glitch are large enough then $\Delta \Phi_{\rm g}$
can quickly exceed a small integer number of rotations. If the photon data are sparse, as is often
the case, this can happen before we have even detected any pulsed photons after the glitch
\citep{Yu2013+Glitches, Clark2015+J1906}.

A result is that the likelihood distribution in the glitch epoch, $t_{\rm g}$, resembles a comb of
possible epochs, with each maximum occurring at a time where $\Delta \Phi_{\rm g}\left(t\right)$
equals an integer number of rotations. Such a highly multi-modal likelihood surface causes problems
for our Monte~Carlo sampling method, as chains can easily get stuck in low-likelihood regions
between maxima, greatly reducing the efficiency of the sampling procedure.

To avoid this, we can include an arbitrary phase increment at the time of the glitch in our phase
model and allow it to vary as part of the Monte~Carlo sampling. This phase increment removes the
multi-modal nature of the likelihood surface by accounting for the phase offset for glitch models
that do not occur at one of the maxima described above. While the phase increment is highly
correlated with the glitch epoch, we can predict and account for the dominant contribution to the
time-dependent part of the glitch increment to remove this correlation, ensuring efficient sampling
(see Appendix \ref{a:glitch_timing} for further details).

Apart from PSR~J1906$+$0722, (published previously in \citet{Clark2015+J1906}) one other pulsar
detected by the \EAH{} survey, PSR~J1844$-$0346, experienced a glitch during the \Fermi{}
mission. Occurring in July--August 2012, with $\Delta f/f \approx 3.5\times10^{-6}$, it was
comparable to some of the largest glitches detected from gamma-ray pulsars
\citep{Pletsch2012+J1838}. With a characteristic age of $\tau_c = 11.6$ kyr, and spin-down energy
$\dot{E} = 4.2\times10^{36}$ erg s$^{-1}$, this pulsar is by far the most energetic pulsar found by
our survey, and also displays

\begin{figure}
  \centering
  \includegraphics[width=\columnwidth]{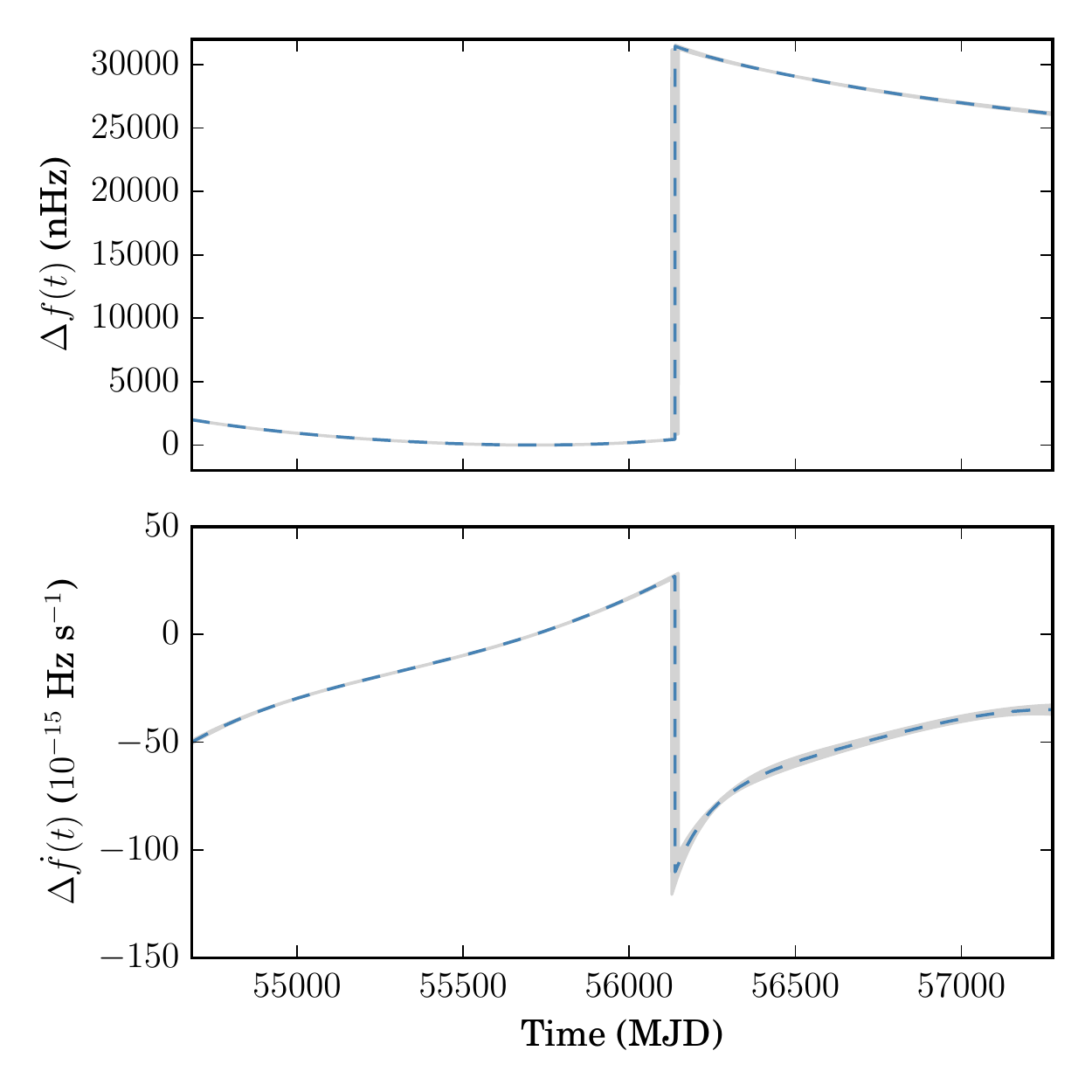}
  \caption{Evolution of the rotational frequency of PSR~J1844$-$0346 during the \Fermi-LAT observation
    time, as in Figure \ref{f:all_freq_evos}.}
  \label{f:J1844_timing}
\end{figure}

\begin{deluxetable}{lc}
  \tablecaption{PSR J1844$-$0346 Glitch Parameters \label{t:glitch_params}}
  \tablecolumns{2}
  \tablewidth{\columnwidth}
  \tablehead{\colhead{Parameter} & \colhead{Value}}
  \startdata
Glitch epoch, $t_{\rm g}$ (MJD) \dotfill & $56135(7)$ \\
Frequency increment, $\Delta f_{\rm g}$ (Hz)\dotfill & $3.06(1)\times10^{-5}$ \\
$\dot{f}$ increment, $\Delta \dot{f}_{\rm g}$ (Hz s$^{-1}$)\dotfill& $-9.4(3)\times10^{-14}$ \\
$\ddot{f}$ increment, $\Delta \ddot{f}_{\rm g}$ (Hz s$^{-2}$)\dotfill & $-7.0(9)\times10^{-22}$ \\
Decaying $f$ increment, $\Delta f_{D,\rm g}$ (Hz)\dotfill& $4.5(7)\times10^{-7}$ \\
Decay time constant, $\tau_{D,\rm g}$ (d)\dotfill& $117(22)$ \\
  \enddata
 \tablecomments{Reported values of glitch parameters are the mean values obtained from the Monte-Carlo analysis described in Section \ref{s:timing} with $1\sigma$ uncertainties in the final digits given in brackets.}
\end{deluxetable}

\noindent a correspondingly large degree of timing noise. The evolution of the
spin frequency and spin-down rate of PSR~J1844$+$0346, including the glitch, are shown in Figure
\ref{f:J1844_timing}. The glitch parameters obtained from the timing analysis are given in Table
\ref{t:glitch_params}.

\subsection{Pulse Profile Variability}
\begin{figure}
  \centering
  \includegraphics[width=\columnwidth]{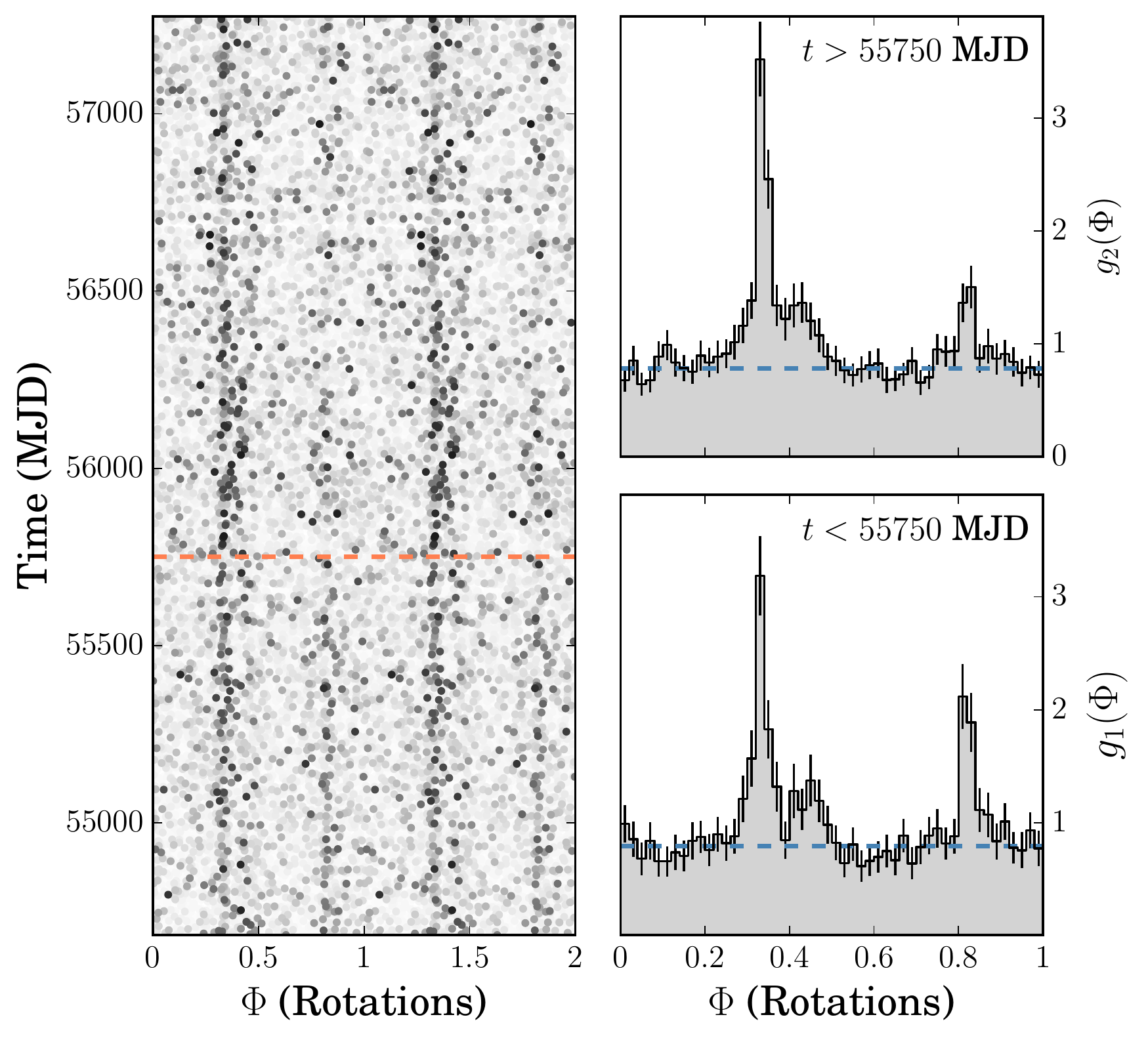}
  \caption{Pulse profile of PSR~J1350$-$6225 showing a possible change around MJD $55750$. Left panel:
    phase--time diagram of the detected photons from PSR~J1350$-$6225, with photon probability weight
    indicated by the intensity. MJD $55750$ is marked by the dashed red line. Right panels, bottom
    and top: normalized phase histograms showing the total pulse profile before ($g_1(\Phi)$) and
    after MJD $55750$ ($g_2(\Phi)$) respectively. The estimated background levels, $b$, are shown by
    the dashed blue line.}
  \label{f:J1350_profiles}
\end{figure}

We checked for variations in the pulse profiles of all pulsars detected in this survey by visually
inspecting their phase--time diagrams, and by measuring their Fourier coefficients in a small number
of time intervals. In one pulsar, PSR~J1350$-$6225, small changes in the first and second Fourier
coefficients were observed. This was also observed in the phase--time diagram for this pulsar (shown
in Figure \ref{f:J1350_profiles}) where the trailing peak seems to appear less prominently in the
latter portion of the \Fermi{} mission than in the earlier data.

To investigate the significance of this effect, we compared the distributions of the observed photon
phases before and after MJD $55750$ (chosen to maximize the change in the pulse profile's first
Fourier coefficient) by calculating the Weighted Kuiper Test statistic
\citep{Jetsu1996+Kuiper},
\begin{equation}
  V_{w_1,w_2} = \max{\left\{G_1(\Phi) - G_2(\Phi)\right\}} + \max{\left\{G_2(\Phi) - G_1(\Phi)\right\}}\,,
\end{equation}
where $G_1(\Phi)$ and $G_2(\Phi)$ are the empirical weighted cumulative distributions of the photon
phases before and after MJD $55750$ respectively.

The distribution of $V_{w_1,w_2}$ under the null hypothesis is unknown, and calculating
significances based on the properties of the unweighted statistic always \textit{under-estimates}
the false-alarm rate. To estimate the significance, we therefore performed a Monte~Carlo
analysis. Using the observed sets of photon weights (before and after MJD $55750$), we randomly
generated two sets of photon phases, with the $j$th photon's phase drawn from a common pulse
profile\footnote{The distribution of $V_{w_1,w_2}$ under the null hypothesis should be independent
  of the chosen pulse profile, since it only tests the possibility that the observed phases are
  drawn from the \textit{same} distribution, regardless of the true underlying distribution.} with
the probability $p = w_j$, otherwise distributed uniformly. This process was repeated many times to
estimate the distribution of $V_{w_1,w_2}$ under the null hypothesis.

From this analysis, we find that the observed value of $V_{w_1,w_2}$ corresponds to a $p$-value of
$0.038$. Given that all 13 pulsars were checked for pulsations, and that a small number of trials
were performed when choosing the date defining the boundary between the two intervals, we conclude
that this is not a significant variation.

Long-term monitoring of the \textit{Fermi}-LAT data from this pulsar would be required to detect the
presence of any pulse profile mode changes, either by observing another such mode change, or by
reducing the uncertainty on the new template pulse profile. A change in pulse profile has only been
detected in one gamma-ray pulsar to date, PSR~J2021$+$4026 \citep{Allafort2013+J2021+4026}. This
variation was accompanied by abrupt changes in the pulsar's gamma-ray flux and spin-down rate,
neither of which are observed from PSR~J1350$-$6225.

\section{Discussion}\label{s:discussion}

\subsection{Sensitivity}\label{s:disc_sensitivity}
Recent publications have argued, both by modeling the emission mechanisms of known radio and
gamma-ray pulsars \citep{Perera2013} and by constructing an unbiased sampling of radio-loud and
radio-quiet gamma-ray pulsars \citep{Rubtsov2015}, that \textit{Fermi}-LAT should detect
significantly more non-recycled gamma-ray pulsars that are radio quiet than are radio loud, by a
factor of $\sim 2$. The 13 new pulsars reported here, only 2 of which appear to be radio loud
(Paper II) bring the total number of radio-quiet and radio-loud non-recycled gamma-ray pulsars to 51
and 61 respectively\footnote{We use the definition from the 2PC that a radio-loud pulsar has a flux
  density $S_{1400} > 30$~$\mu$Jy. Two gamma-ray pulsars have radio detections with lower fluxes, we
  count them here as radio-quiet.}. This would suggest that there are still large numbers of
unidentified radio-quiet gamma-ray pulsars requiring blind gamma-ray pulsation searches to be
detected. In this section, we compare the newly discovered pulsars to the earlier population of
gamma-ray pulsars discovered in blind searches to identify and discuss the overall trends in blind search
sensitivity.

One question that we may like to address is, how bright does an unidentified gamma-ray pulsar need
to be in order to be detectable in a blind search? In particular, we would like to know the lowest
point-source significance within which pulsations can be detected, and how this threshold changes as
more data are accumulated.

\begin{figure}
  \centering
  \includegraphics[width=\columnwidth]{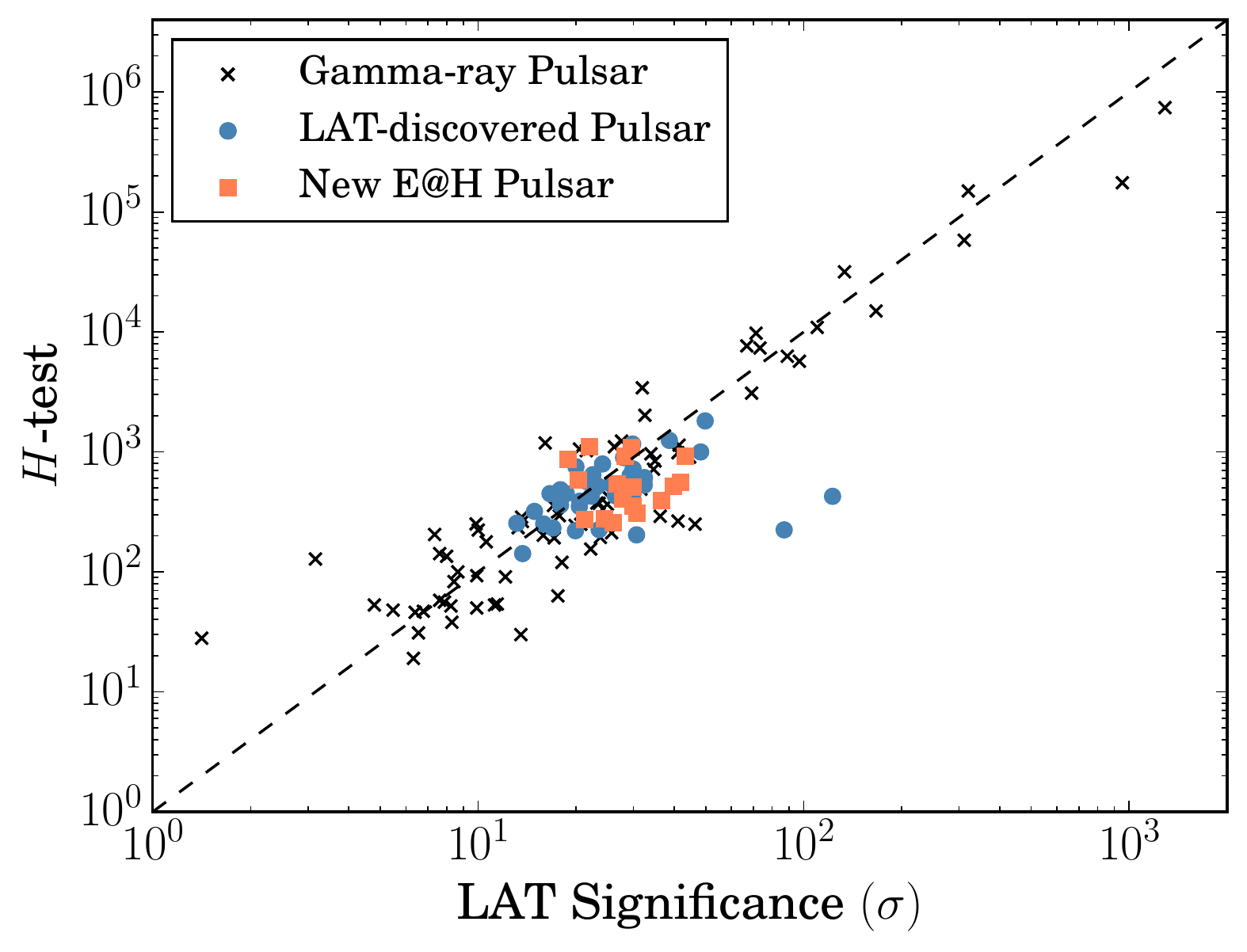}
  \caption{Point-source significance vs. $H$-test value for all pulsars in 2PC, and those detected
    by the \EAH{} survey. The dashed line denotes where $H = \rm{TS}$. The $H$-test values and
    significances for previously detected blind-search pulsars have been scaled back from their 2PC
    values to estimate their value in the data used in their original searches.}
  \label{f:pulsar_significances}
\end{figure}

In Figure \ref{f:pulsar_significances}, we have plotted the point-source significance versus the $H$-test
value for each pulsar in the 2PC, as well as the blind search pulsars detected after this catalog
was produced \citep{Pletsch+2013-4pulsars}, and those discovered by this survey \citep[including
  PSR~J1906$+$0722,][]{Clark2015+J1906}. We can see that the $H$-test value for a pulsar can be well
approximated by its point-source Test Statistic (TS) value (as shown by the dashed line in Figure
\ref{f:pulsar_significances}). The values plotted here for pulsars detected in previous blind
searches have been scaled back to represent their value at the time of their discovery.

We begin by looking at how the coherent detection statistic, $\mathcal{P}_1$, varies with the
observation duration, $T_{\rm obs}$. Since the $H$-test is a maximized sum over $\mathcal{P}_n$
values, the relevant scalings with respect to the observation time will be unchanged. As shown in
\citet{Methods2014}, the expected values of $\mathcal{P}_1$ for a signal with average photon arrival
rate, $\mu$, pulse profile Fourier coefficients, $\left\{\gamma_n\right\}$, and pulsed fraction
$p_{\rm s}$ over an observation lasting $T_{\rm obs}$ is approximately
\begin{equation}
  E_p\left[\mathcal{P}_1\right] \approx 2 p_{\rm s}^2 \,\mu \, \left|\gamma_1\right|^2 \, T_{\rm
    obs} + 2\,.
  \label{e:coh_SN}
\end{equation}
This shows the well-known result that $\mathcal{P}_1$ (and hence $H$) increases linearly with
time. This is relevant for detecting gamma-ray pulsations using known radio or X-ray ephemerides;
since only a small number of trials are required, fully coherent searches are perfectly feasible and
signals only need to overcome a low threshold to be detected. As $T_{\rm obs}$ increases, so too
does the point-source significance, and pulsars whose gamma-ray pulsations are not initially above
the detection threshold will eventually be detectable. Indeed, pulsations have been detected in this
way from sources all the way down to the point-source detection threshold \citep{Hou2014+FaintPSRs}.

However, the limiting factor in our blind searches is the sensitivity of the initial semicoherent
stage. The expected semicoherent (power) \ac{S/N}, given a lag-window length, $T$, is
\begin{equation}
  \theta^2_{S_1} = E_p\left[S_1\right] \approx p_{\rm s}^2 \,\mu \, \left|\gamma_1\right|^2 \, T^{1/2} \, T_{\rm
    obs}^{1/2}\nonumber\,.
  \label{e:scoh_SN}
\end{equation}
The semicoherent S/N accumulates much more slowly, only with the square root of $T_{\rm
  obs}$. Substituting this into Equation (\ref{e:coh_SN}), we can identify the effective coherent
threshold, $\mathcal{P}_1^{\ast}$, in terms of the semicoherent S/N threshold, $S_1^{\ast}$, as
\begin{equation}
  \mathcal{P}_1^{\ast} \approx 2\, S_1^{\ast} \left(\frac{T_{\rm
      obs}}{T}\right)^{\frac{1}{2}} + 2\,.
  \label{e:P1_thresh}
\end{equation}
This reveals an unintuitive result: with a fixed semicoherent threshold and lag-window size, as the
observation time increases our sensitivity threshold in terms of the coherent signal power (and
hence point-source significance) actually \textit{increases}.

Equation (\ref{e:P1_thresh}) also reveals the solution to this problem: if we are to maintain the
same source significance threshold in searches using longer observation times, we must also increase the
lag-window duration by the same factor. However, as was derived in \citet{Methods2014}, the
computational cost associated with a blind semicoherent search scales with $T^4 \, T_{\rm
  obs}$.

\begin{figure}
  \centering
  \includegraphics[width=\columnwidth]{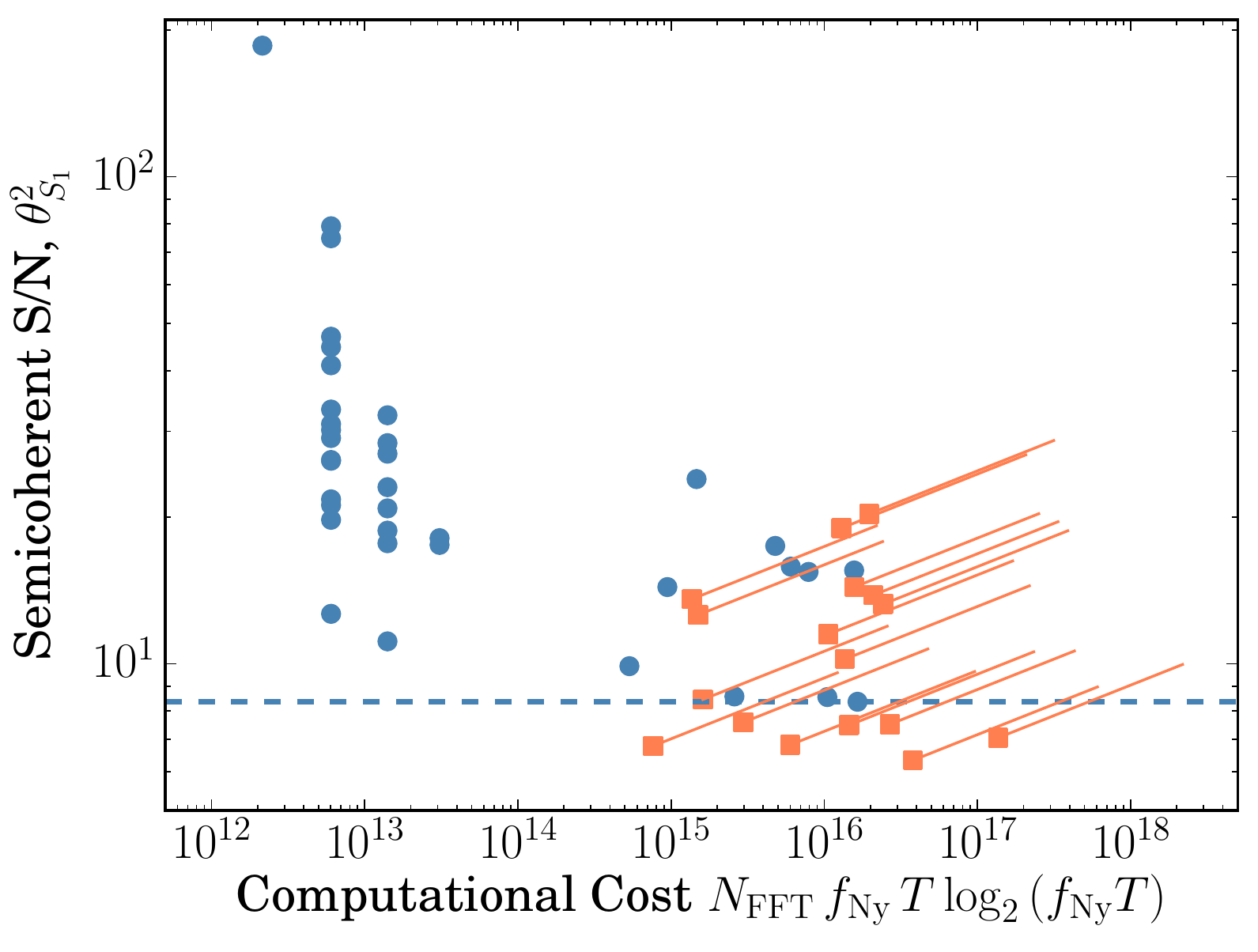}
  \caption{Computing cost vs. Semicoherent signal-to-noise ratio comparison for all LAT-discovered
    gamma-ray pulsars. Those newly discovered in this work are shown by orange squares, previously
    detected pulsars are shown as blue circles. The locations on this plot of the newly detected
    pulsars represent their expected location if our survey had used the 50\% shorter lag-window
    used in previous searches. The lowest semicoherent S/N for all blind-search pulsars is given by
    the blue dashed line, we consider signals below this line to be undetectable. The orange
    diagonal lines show how the computational cost and S/N increased for the newly discovered
    pulsars due to the longer lag window used in this survey. }
  \label{f:sens_compcost}
\end{figure}

In Figure \ref{f:sens_compcost}, we have estimated the semicoherent S/N at the time of discovery for
all gamma-ray pulsars detected in previous blind searches by calculating their $\mathcal{P}_1$
values from the data provided by the 2PC, and scaling these down to the $T_{\rm obs}$ used in each
search. We have also estimated the computational cost that would be required to perform each search
(only covering the young pulsar parameter space) using the original lag-window size ($T = 2^{20}$~s)
and observation length, but with otherwise the same search scheme described in Section
\ref{s:scheme}. Searches prior to \citet{Pletsch+2012-9pulsars} only searched for pulsations from
the LAT point-source location, rather than searching over many possible sky locations. While this
significantly decreases the required computational cost, this requires pulsars either to be close to
the LAT source's estimated position, or to have a strong pulsed signal, such that they can be
detected despite a large positional offset. This also rules out the detection of isolated
MSPs, whose high frequency requires us to search sky locations with much finer
resolution. We therefore exclude the cost of searching for MSPs from our estimated
computing costs.

In this figure, we have also estimated the computational cost and S/N that each of the new pulsars would
have had if we had performed these searches with the lag-window duration used by previous
searches. At least four, possibly as many as seven pulsars would have been below the lowest semicoherent
S/N from a previously detected pulsar, and therefore most likely would not have been detected had we not
performed the more expensive search with the longer lag window.

Figure \ref{f:sens_compcost} also shows how the computing cost and S/N increased for these pulsars
as a result of using a lag-window with double the length. We see that only one pulsar detected here
required less computing time than any previously detected gamma-ray pulsar, with the most expensive
detection being more than $100$ times more costly. Based on this argument, the current survey should
have taken many years to complete, even taking into account a generous estimate of a doubling of the
computing power between the two surveys computed on \EAH{}. In actuality, the first 118 sources were
searched in just over one year, similar in duration to the first gamma-ray pulsar survey. We
consider this a testament to the improved efficiency of the search methods developed in
\citet{Methods2014}.

The signal detected here with the lowest coherent power was from PSR~J1105$-$6037 with
$\mathcal{P}_1 = 165$. From the spread about the diagonal of Figure \ref{f:pulsar_significances}, we
estimate that a pulsar with a particularly narrow pulse profile should be detectable with a TS of
around $100$ (a significance of approximately $10\sigma$).  The lowest-significance point source in which a
pulsar was detected here (PSR J1350$-$6225) had a TS of $356.6$ (approximate significance $19\sigma$). This
could be because the ranking method for our target sources depends on the sources' significance;
brighter \Fermi-LAT sources were searched earlier than dimmer sources with similar spectral
properties. It is therefore possible that as our survey continues we may yet detect a pulsar with a
lower point-source significance.

In any case, the following trend is clear: as the \textit{Fermi} mission continues, dimmer pulsars
will gradually become detectable to blind searches, but the computing cost required to detect
equally low $H$-test signals will rise quickly as we search longer data sets. This survey received a
large boost to its sensitivity in the form of the ``free'' (in terms of computational cost) increase
in the observed photon counts offered by the Pass 8 data reprocessing \citep{Pass8}. It is unlikely
that such a welcome jump in sensitivity will occur again in the near future, and therefore it may
become necessary to make some sacrifices to maintain the current search sensitivity without
the computing cost requirements becoming unrealistic. For example, reducing the spin-down range to
which we are sensitive by a factor of $\sim 10$ will reduce the computational cost by the same
factor, while only losing sensitivity to the most energetic of pulsars. Indeed, one could argue that
we already have greatly reduced sensitivity to these pulsars since they typically exhibit the most
timing noise and large glitches.

Another option would be to focus our available computing resources to perform more sensitive
searches of only the most promising target sources, rather than performing wide-ranging surveys
of many unidentified sources as was done here. Indeed, if we look at how the computing cost was
distributed amongst sources, as shown in Figure \ref{f:ps_nffts}, we can see that, even with a
favorable pulse profile, some sources would require a pulsed fraction far higher than unity for us
to detect pulsations. The procedures described in Section \ref{s:sensitivity} allow us to reliably
predict our search sensitivity to new sources for the first time. Now that we are able to predict
this in advance, we can remove such sources from our search list and focus our efforts on improving
our sensitivity to the more promising sources. If we relax the requirement of 95\% detection
probability used in producing Figure \ref{f:ps_nffts} slightly to reflect a more moderate chance of
detecting pulsations from each source, we find that $\sim25\%$ of the computing cost of this survey
could perhaps have been better spent by removing unpromising sources.

\subsection{New Pulsars}\label{s:disc_pulsars}
Only energetic pulsars with spin-down powers above $10^{33}$\,erg\,s$^{-1}$ are observed to emit
gamma-ray pulsations \citep{Guillemot2016+Deathline}. The group of non-recycled gamma-ray pulsars
therefore lies at the young and energetic end of the overall pulsar population, with typical
characteristic ages less than a few million years, and surface magnetic field strengths between
$10^{11}$--$10^{13}$\,G. This contrasts with the population of non-recycled radio pulsars
(which extends to characteristic ages of up to 1 Gyr, and spin-down powers as low as
  $10^{30}$\,erg\,s$^{-1}$, \citealt{ATNF2005}) and with magnetars (whose surface magnetic fields are
  in excess of $10^{14}$\,G, \citealt{Turolla2015+Magnetars}). In these respects, the group of pulsars
detected here appears to be consistent with the overall population of young gamma-ray pulsars (see
Figure \ref{f:f_fdot_diagram}, and Table \ref{t:params2}).

Two of the new pulsars reported here (PSR~J1057$-$5851 and PSR~J1827$-$1446) are the most slowly
spinning gamma-ray pulsars yet discovered, at $1.6$~Hz and $2.0$~Hz respectively. While the
existence of these pulsars is not in tension with estimates of the gamma-ray pulsar ``death line''
\citep{Wang2011+DeathLine}, their discovery does extend the known population of gamma-ray pulsars
down to lower spin frequencies. Sensitive blind searches that can detect faint young pulsars are
important to fully explore the low-$\dot{E}$ region of the gamma-ray pulsar population, and to
reduce biases inherent in it \citep{Guillemot2016+Deathline}.

Also of interest are the pulsars detected here despite exhibiting significant timing noise, PSR
J1844$-$0346, and PSR~J0359$+$5414. The large contributions of higher frequency derivatives mean
that the original phase model used in the blind search could not maintain phase coherence over the
full duration of the data. These effects make it very difficult for the coherent follow-up stages to
pick up these signals. On the other hand, the semicoherent stage is largely unaffected by timing
noise, although large glitches are also detrimental to the semicoherent sensitivity. Noisy pulsars
will therefore only appear as semicoherent candidates, and may easily escape detection from our
pipeline, which focuses on the results from the final coherent follow-up. Further investigation of
the vast number of semicoherent candidates reported by \EAH{} may yet reveal more noisy pulsars
lurking in our results.

\section{Conclusions}\label{s:concl}
We have presented the discovery of 13 new gamma-ray pulsars found by the ongoing
\EAH{} survey of unidentified \textit{Fermi}-LAT sources. Amongst these pulsars are
two new energetic pulsars with $\dot{E} > 10^{36}$ erg s$^{-1}$, one of which experienced a large
glitch; and the two slowest spinning gamma-ray pulsars yet detected.

As the \textit{Fermi} mission continues and the LAT gathers more data, the sensitivity to weak
pulsar signals will increase, and many currently undetectable pulsars could rise above the detection
threshold in the near future, although future searches with more data will also require even more
computing power to be sensitive to similarly weak signals.

We also placed realistic upper limits on the pulsed flux from point sources from which no pulsations
were detected. The framework for this allows us to also predict our sensitivity to other sources,
enabling us to identify promising targets for searching, and also to veto sources from which
pulsations would be almost impossible to detect. This will allow us to focus our computing power on
increasing our sensitivity to the most promising sources in future surveys.

A further exciting new advancement is the recent launch of the first \EAH{} survey for gamma-ray
pulsars in candidate binary systems with well-constrained orbital parameters, similar to the search
that discovered PSR~J1311$-$3430 \citep{Pletsch+2012-J1311}. The additional computing power of
\EAH{} will enable more complicated searches, allowing for searches from sources with larger
uncertainties in their orbital parameters, or even with slight eccentricities.

\acknowledgements

We are extremely grateful to all volunteers who have donated their CPU time to the
\textit{Einstein@Home} project, without whom this survey could not have been performed. We are
especially grateful to those users whose computers discovered the 13 new pulsars reported here. They
are\footnote{Where the volunteer's name is unknown or private, we give the \textit{Einstein@Home}
  username in quotation marks.}:

\begin{itemize}
\item{PSR~J0002$+$6216: James Drews of UW-Madison, WI, USA and Ralph Elwell of Richland, WA, USA;}
\item{PSR~J0359$+$5414: Whelton A. Miller III, Lincoln University of Pennsylvania \& University of
  Pennsylvania, USA; the ATLAS Cluster, AEI, Hannover, Germany and Philip ``Delty'' Horney of the
  GPU Users Group, Fort Wright, KY, USA;}
\item{PSR~J0631$+$0646: Katagiri, Atsushi of Kawasaki, Japan and Nicholas Huwar of Houston, TX, USA;}
\item{PSR~J1057$-$5851: Syracuse University HTC Campus Grid\footnote{Supported by NSF awards ACI-1341006
  and ACI-1541396}\addtocounter{footnote}{-1}\addtocounter{Hfootnote}{-1}, NY, USA; Igor Yakushin of
  Chicago, IL, USA and the LIGO Laboratory, USA}
\item{PSR~J1105$-$6037: The ATLAS Cluster, AEI, Hannover, Germany and Syracuse University HTC Campus Grid\footnotemark\addtocounter{footnote}{-1}\addtocounter{Hfootnote}{-1}, NY, USA;}
\item{PSR~J1350$-$6225: Petr Ruzicka of Brno, Czech Republic and Bryden Kanngiesser of Calgary, Canada;}
\item{PSR~J1528$-$5838: ``fred c'' and Gabriel Vasquez of Miami, FL, USA;}
\item{PSR~J1623$-$5005: Lars Bollwinkel, of Kiel, Germany and Greg Dorais of Martinez, CA, USA;}
\item{PSR~J1624$-$4041: Xio of NYC and Hung Tran of Chandler, AZ, USA;}
\item{PSR~J1650$-$4601: Syracuse University HTC Campus Grid\footnotemark, NY, USA and Eric Schwartz
  of Vashon Island, WA, USA;}
\item{PSR~J1827$-$1446: The ATLAS Cluster, AEI, Hannover, Germany; Igor Yakushin of
  Chicago, IL, USA and the LIGO Laboratory, USA}
\item{PSR~J1844$-$0346: Aur\'elien FAUCHEUX of Antibes, France and Roger Gulbranson, Ph.D. of Wickliffe,
  OH, USA;}
\item{PSR~J2017$+$3625: Kurt Gramoll, Ph.D., University of Oklahoma, OK, USA and Michael Brandau, of
  Kassel, Germany.}
\end{itemize}

This work was supported by the Max-Planck-Gesell\-schaft~(MPG), by the Deutsche
Forschungsgemeinschaft~(DFG) through an Emmy Noether research grant PL~710/1-1 (PI:
Holger~J.~Pletsch), and by NSF award 1104902.

The \textit{Fermi} LAT Collaboration acknowledges generous ongoing support from a number of agencies
and institutes that have supported both the development and the operation of the LAT as well as
scientific data analysis.  These include the National Aeronautics and Space Administration and the
Department of Energy in the United States, the Commissariat \`a l'Energie Atomique and the Centre
National de la Recherche Scientifique / Institut National de Physique Nucl\'eaire et de Physique des
Particules in France, the Agenzia Spaziale Italiana and the Istituto Nazionale di Fisica Nucleare in
Italy, the Ministry of Education, Culture, Sports, Science and Technology (MEXT), High Energy
Accelerator Research Organization (KEK) and Japan Aerospace Exploration Agency (JAXA) in Japan, and
the K.~A.~Wallenberg Foundation, the Swedish Research Council and the Swedish National Space Board
in Sweden.

Additional support for science analysis during the operations phase is gratefully acknowledged from
the Istituto Nazionale di Astrofisica in Italy and the Centre National d'\'Etudes Spatiales in
France.
\appendix

\section{Candidate Ranking}
\label{a:PFA_ranking}
In the final follow-up stages, performed offline, we would like to only search the most significant
candidates. To rank candidates by their significance, we need to account for the effective
number of trials from which each candidate has resulted. While the number of (semicoherent)
independent trials is approximately the same in each work unit within each source, there are more
work units in higher frequency bands due to the density of sky locations increasing with
frequency. Additionally, since the grid of sky locations searched in the first stage is constructed
at the highest frequency in the band, whereas the ``zoomed in'' grids are defined by the candidates'
spin frequencies, the number of trials in the refinement step varies from candidate to candidate.

The overall result of these effects is that candidates with high detection statistic values are more
likely to occur by chance in higher frequency bands than in lower frequency bands, and at the higher
end of the frequency band. We construct a consistent ranking statistic by comparing candidates'
false alarm probabilities whilst taking the differing number of trials into account.

We start from the result \citep{Kruger2002} that the \ac{cdf}, $G(X_{\rm max})$, of the maximum value of $N$
samples of the random variable $X$, is related to $F(X)$, the \ac{cdf} of $X$, by
\begin{equation}
G(X_{\rm max}) = \left[F(X_{\rm max})\right]^N\,.
\end{equation}
The false alarm probability of $X_{\rm max}$ after $N$ samples is therefore
\begin{align}
P_{\rm{FA},N}(X_{\rm max}) &= 1 - \left[1 - P_{{\rm FA},1}(X_{\rm max}) \right]^N\nonumber\\
&\approx N P_{{\rm FA},1}(X_{\rm max})\,,
\end{align}
where we have assumed that $P_{{\rm FA},1} \ll 1/N$. 
In our case, the single-trial false alarm for a candidate with coherent power $\mathcal{P}_1$ is
\begin{align}
P_{{\rm FA},1}(\mathcal{P}_1) &= \int_{\mathcal{P}_1}^{\infty}
\chi_2^2\left(\mathcal{P}_1^\prime\right)\, d\mathcal{P}_1^\prime\\ &= e^{-\mathcal{P}_1/2}\;,
\end{align}
where $\chi_2^2(X)$ is the central chi-squared distribution with two degrees of freedom. 

It is considerably more difficult to estimate the effective number of independent trials, since each
candidate is the result of a large number of trials in previous search stages using different
detection statistics. However, since at this stage we are only interested in ranking candidates
within each source, and the number of independent trials in the semicoherent step is approximately
the same for each candidate, we only need to consider the varying number of trials in the follow-up
stages, and the total number of work units in each frequency band.

The overall false-alarm probability is therefore a function of the frequency of the candidate, $f$,
and the coherent power:
\begin{equation}
P_{\rm FA}(\mathcal{P}_1, f) = K N_{\rm W}(f) \, N_{\rm F}(f) \, e^{-\mathcal{P}_1/2}\,,
\end{equation}
where the constant of proportionality, $K$, is the (unknown) number of independent trials per work
unit, $N_W(f)$ is the number of work units within the appropriate frequency band and $N_F(f)$ is the
number of trials in the coherent follow-up stage for a candidate at frequency $f$.

We define the ranking statistic, $\hat{R}$ for follow-up analyses according to the logarithm of the
inverse of the false alarm probability,
\begin{align}
\hat{R}(\mathcal{P}_1,f) &\equiv -\log \left[\frac{P_{\rm FA}(\mathcal{P}_1, f)}{K}\right] \nonumber \\
&= \frac{\mathcal{P}_1}{2} -\log\left[N_W(f)\right] -\log\left[N_F(f)\right]\,,
\end{align}
where we have removed the constant term corresponding to $K$.  We note that the above formulation of
$\hat{R}$ can \textit{only} be used to rank pulsar candidates from the same source, as the effective
number of independent trials per work unit ($K$) varies from source to source.

\section{Distribution of $S_1$ with a Signal}
\label{a:app_signal_stats}
To estimate the sensitivity of the search, it is necessary to know the expected distribution of
$\hat{S_1}$ in the presence of a signal for each source. To derive this, we first expand the double sum
of Equation (\ref{e:s1_definition}) and separate it into terms in which photon indices are never
equal. Identifying the photon weights as the probability that each photon originated from the source
in question, and therefore assuming that each photon has a probability $w_j p_{\rm s}$ of having
been pulsed, we find the following expressions,
\begin{subequations}
\begin{align}
  E_p\left[w_j^m e^{-in\Phi(t_j)}\right] &= w_j^{m+1} \, p_{\rm s} \, \gamma_n \,,\\
  E_p\left[w_j^m e^{in\Phi(t_j)}\right] &= w_j^{m+1} \, p_{\rm s} \, \gamma_n^{\ast}\,.
\end{align}
\end{subequations}
The expectation value and variance of $S_1$ in the presence of a signal are therefore found, after
some relabeling and rearranging, to be
\begin{equation}
  E_p(p_{\rm s},\left\{\gamma_n\right\})[S_1] = \frac{p_{\rm s}^2 \left| \gamma_1 \right|^2}{\kappa_{S_1}} \sum_{j=1}^N \sum_{k\neq j}^{N}
  w_j^2 w_k^2 \hat{W}^{\rm rect}_T(\tau_{jk})\,,
\end{equation}
\begin{multline}
  \sigma_p^2(p_{\rm s},\left\{\gamma_n\right\})[S_1] = \frac{1}{\kappa_{S_1}^2}\sum_{j=1}^N \sum_{k\neq j}^{N} \left[ w_j^2 \, w_k^2 \, \hat{W}^{\rm
    rect}_T(\tau_{jk}) \left( \vphantom{\sum_{l\neq j
        \neq k}^{N}}1 + p_{\rm s}^2 \left| \gamma_2 \right|^2 w_j \, w_k \, - 2 p_{\rm s}^4
    \left| \gamma_1 \right|^4  w_j^2 \, w_k^2 \right. \right. \\
    + \left. \left. \sum_{l\neq j \neq k}^{N} w_l^2 \, \hat{W}_T^{\rm rect}(\tau_{jl}) \left[ 2 p_{\rm s}^2 \left| \gamma_1
      \right|^2 + 2 p_{\rm s}^3 \Re\left(\gamma_2 \left(\gamma_1^{\ast}\right)^2\right) w_j  - 4 p_{\rm s}^4 \left|
      \gamma_1 \right|^4 w_j^2 \right] \right) \right]\,.
\end{multline}
The expected semicoherent \ac{S/N} for a signal with pulsed fraction $p_{\rm s}$ and a pulse profile with
Fourier coefficients $\left\{\gamma_n\right\}$ is therefore
\begin{align}
  \theta_{S_1}^2(p_{\rm s},\left\{\gamma_n\right\}) &= \frac{E_p[S_1] - E_0[S_1]}{\sqrt{\sigma_0^2[S_1]}} \nonumber \\
  &= p_{\rm s}^2
  \left|\gamma_1\right|^2 \sqrt{\sum_{j=1}^N \sum_{k \neq j}^{N} w_j^2 w_k^2 \, \hat{W}^{\rm
      rect}_T(\tau_{jk})}\,.
\end{align}
In addition to the statistical variance of $S_1$, a signal at a random location in the parameter
space will be detected at the nearest grid point, and some signal power will be lost as a result of
this offset. Denoting this mismatch by $m$, the \ac{pdf} of $\hat{S_1}$ is therefore the \ac{pdf} of
the product of $S_1$ and $(1 - m)$, which we approximate as a Gaussian with the same mean and variance,
\begin{subequations}
  \begin{align}
    E_p\left[\hat{S_1}\right] &= E_p\left[S_1 \left(1 - m\right)\right] = E_p\left[S_1\right] (1 - E\left[m\right])\,,\\
    \sigma^2_p\left[\hat{S_1}\right] &= \sigma_p^2\left[S_1\right]\, \sigma^2\left[m\right] + \sigma_p^2\left[S_1\right]\,\left(1 - E\left[m\right]\right)^2 +
    \sigma^2\left[m\right] \,E_p\left[S_1\right]^2\,.
  \end{align}
\end{subequations}
The values of $E\left[m\right]$ and $\sigma^2\left[m\right]$ depend on the geometry of the search grid. For
the grid used in this survey, which had a maximum mismatch per parameter of $0.15$ and an interbinned
frequency spectrum, $E\left[m\right] \approx 0.22$ and $\sigma^2\left[m\right] \approx 8\times10^{-3}$.
This \ac{pdf} is illustrated in Figure \ref{f:detection_probability}.

\section{Efficient Sampling of Glitch Parameters}\label{a:glitch_timing}
As mentioned in Section \ref{s:J1844}, when timing large glitches, a phase increment occurring at
the time of the glitch can be included in the phase model to ensure that the likelihood surface in
the glitch epoch is continuous and easy to sample.

\begin{figure}
    \centering
    \includegraphics[width=\columnwidth]{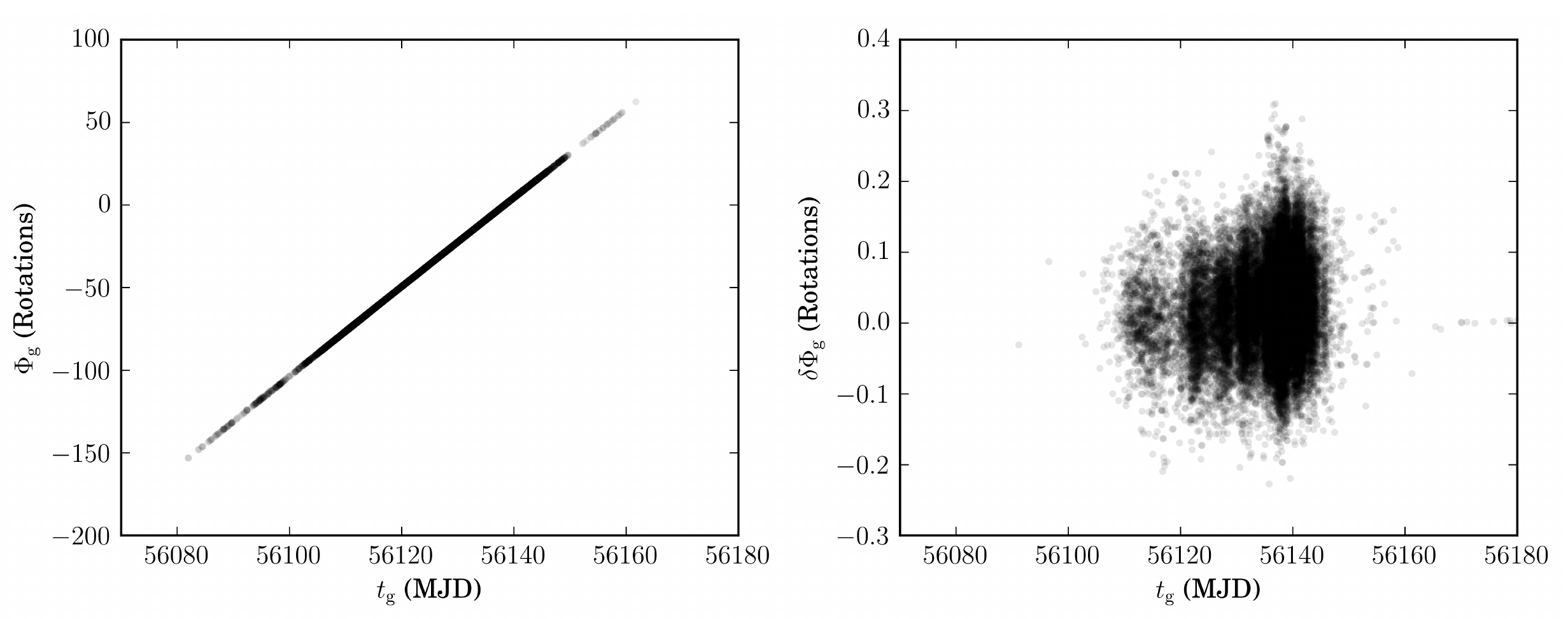}
    \caption{Glitch epochs and phase increments from Monte~Carlo sampling during timing of PSR
      J1844$-$0346. In the left panel, the phase increment was completely free to vary and exhibits a
      high correlation with the glitch epoch. In the right panel, the glitch epoch-dependent part of
      the phase increment was accounted for, and a small random increment was allowed to vary on top of
      this, removing the correlation.}
  \label{f:glitch_correlation}
\end{figure}

However, this increment is strongly correlated with the glitch epoch, as can be seen in the first
panel of Figure \ref{f:glitch_correlation}, which can lead to inefficient
Monte~Carlo sampling. Once an initial combination of valid glitch parameters has been found
(including a suitable combination of glitch epoch and phase increment that removes any phase
discontinuity at the glitch) we can remove this correlation by separating the total phase increment
in our phase model into separate terms: a glitch epoch-dependent term accounting for the known
difference in phase between the glitch model parameters being sampled and the initial ``reference''
glitch model, plus a random offset that is allowed to vary as part of the
Monte~Carlo sampling.  Denoting the reference glitch model with the subscript ${\rm g_0}$, and the
sampled glitch model parameters by the subscript ${\rm g_1}$, the total phase increment is
\begin{equation}
  \Phi_{\rm g_1} = \delta \Phi_{\rm g_1} + \Phi_{\rm g_0} +
  \begin{cases}
    -\Delta \Phi_{\rm g_1} \left(t_{\rm g_0}\right), & t_{\rm g_1} < t_{\rm g_0}\\
    \hphantom{-}\Delta \Phi_{\rm g_0} \left(t_{\rm g_1}\right), & t_{\rm g_1} > t_{\rm g_0}
  \end{cases}\,,
  \label{e:glitch_phase1}
\end{equation}
where, for $i = 0,1$,
\begin{equation}
  \Delta \Phi_{\rm g_i}(t) = 2\pi\,\left[ \Delta f_{\rm g_i}(t - t_{\rm g_i}) + \frac{\Delta
    \dot{f}_{\rm g_i}}{2}(t - t_{\rm g_i})^2 + \frac{\Delta \ddot{f}_{\rm g_i}}{6} (t - t_{\rm g_i})^3 +
  \Delta f_{\rm D, g_i} \tau_{\rm D,g_i} \left( 1 - {\rm e}^{-(t - t_{\rm g_i})/\tau_{\rm D,g}}\right)\right]\,.
\end{equation}

The first term of Equation (\ref{e:glitch_phase1}) ensures that the phase increment is free to vary
over a small range to find the glitch parameters with the highest likelihood. The other terms
ensure that the reference glitch model's desirable property of causing no large phase discontinuity
between the pulse before and after the glitch also applies to the sampled glitch parameters.  This
ensures that the sampling rate (and hence the efficiency of the timing procedure) is not
unnecessarily burdened by having to phase-fold the data with glitch models resulting in large phase
discontinuities, which will obviously have a low likelihood and be rejected.

Removing the correlation between the glitch epoch and the part of the phase increment that is being
sampled ensures that the Monte~Carlo chains explore the parameter space efficiently. This is
especially helpful at the beginning of the Monte~Carlo run, as the starting locations of the chain
are spread uniformly throughout the parameter space, and could otherwise easily get stuck in
low-likelihood regions, as they struggle to jump to the very narrow, highly correlated region of
high likelihood.

\listofchanges

\end{document}